\documentclass{jpp}
\usepackage{graphicx}

\usepackage[utf8]{inputenc}
\usepackage[T1]{fontenc}
\usepackage{amsmath}
\usepackage{multirow}
\usepackage{xcolor}

\shorttitle{Collisionless WHFI}
\shortauthor{R. Davies, P. P. Choudhury, and A. F. A. Bott}

\title{Collisionless whistler heat-flux instability in ultra-high-$\beta$ plasmas}

\author{Rhisiart Davies\aff{1}
  \corresp{\email{rhisiart.davies@physics.ox.ac.uk}},
  Prakriti Pal Choudhury\aff{1}
 \and Archie F. A. Bott\aff{1}}

\affiliation{\aff{1}Department of Physics, University of Oxford, Parks Road, Oxford, OX1 3PU, UK}

\begin{document}

\maketitle

\begin{abstract}
Kinetic instabilities, notably the whistler heat-flux instability (WHFI), are known to suppress thermal transport significantly in the moderate- to high-$\beta$ plasmas relevant to many astrophysical systems. This paper explores WHFI-regulated heat transport in a new regime: ultra-high-$\beta$ plasmas with $\beta_{e} \gtrsim L_{\mathrm{T}}/\rho_e$ (where $\beta_e$ is the ratio of electron thermal to magnetic pressure, $\rho_{e}$ is the electron Larmor radius, and $L_{\mathrm{T}}$ is the macroscopic temperature-gradient length scale). Extrapolating previous theories of the WHFI to ultra-high-$\beta$ plasmas, we propose that the magnetic energy in unstable whistler fluctuations becomes comparable to that of the background magnetic field at saturation. We corroborate this hypothesis using 1D3V and 2D3V kinetic simulations using the particle-in-cell code OSIRIS. We also find that, in ultra-high-$\beta$ plasmas, the heat flux is localised and no longer regulated primarily by resonant pitch-angle scattering of electrons; instead, thermal energy is transported predominantly by advection at the phase velocity of energetically dominant whistlers. 
Heat-flux suppression is observed in both 1D3V and 2D3V simulations; however, we show that the saturation of the WHFI and the regulation of heat flux are sensitive to dimensionality in the ultra-high-$\beta$ regime. Both the amplitude and phase velocity of the heat-flux-regulating whistler waves scale differently with $\beta_e$, yielding parallel heat fluxes, normalised to the free-streaming value, of $q_{e\parallel} / q_\mathrm{fs} \approx 4.7 \beta_{e}^{-1}$ and $q_{e\parallel} / q_\mathrm{fs} \approx 0.3 \beta_{e}^{-1/2}$ in 2D3V and 1D3V simulations, respectively. We also perform 2D3V simulations with background magnetic fields inclined to the temperature gradient, showing that perpendicular heat transport remains negligible in ultra-high-$\beta$ plasmas; this indicates that anomalous diffusive processes contribute negligibly to cross-field transport. We develop a heuristic theory from kinetic equations that explains these phenomena. Our work extends our understanding of how the WHFI modifies thermal transport to regimes applicable to high-energy-density physics and the reionised intergalactic medium. 
\end{abstract}

\section{Introduction}

Understanding the transport properties of plasmas in various physical regimes has long attracted the attention of plasma physicists. A key achievement of twentieth-century plasma physics was the derivation of transport models in classical, collisional plasmas using kinetic theory. Such plasmas are weakly coupled ($n_e \lambda_{\rm D}^3 \gg 1$, where $n_e$ is the number density of electrons and $\lambda_{\rm D}$ is the Debye length), non-degenerate ($n_e \lambda_{\mathrm{th}e}^{3} \gg 1$, where $\lambda_{\mathrm{th}e}$ is the electron thermal de Broglie wavelength) and have thermal electron mean free paths $\lambda_e$ that are much smaller than the macroscopic length scale 
$L_\mathrm{T} = (-\nabla_\parallel\ln{T_e})^{-1}$ over which the electron temperature $T_e$ varies. For such plasmas, \citet{spitzer_transport_1953} developed a model in which the electron heat flux is given (in a hydrogen plasma) by $q_e = -3.16(n_e v_{\mathrm{th}e} T_e) \lambda_e / L_\mathrm{T}$, where $v_{\mathrm{th}e}$ is the electron thermal velocity. For sufficiently large values of $\lambda_e / L_\mathrm{T}$, the heat flux no longer follows the Spitzer-H\"arm model, but is instead limited to some fraction $\alpha_0$ of the so-called free-streaming value: $q_e \leq \alpha_0 q_\mathrm{fs}$ \citep{malone_indications_1975, bell_elecron_1981, li_study_2010}, where $q_\mathrm{fs} \sim n_e v_{\mathrm{th}e} T_e$ (an exact definition is given in (\ref{eq:qfs_defn})). A magnetic field of sufficient strength -- that is, one that causes the electron Larmor radius $\rho_e$ to fall below the electron mean free path $\lambda_e$  -- introduces anisotropy into the system by limiting transport across magnetic field lines, as has been described by \citet{braginskii_transport_1965}. In such a system, the heat transport perpendicular to the magnetic field is suppressed by a factor $(\rho_e/\lambda_e)^{2}$ compared to the Spitzer-H\"arm model, whilst the parallel heat flux remains unchanged.

For many decades, it has been suspected that these classical estimates of the parallel heat flux fail in weakly collisional and collisionless, high-$\beta$ plasmas  ($\beta \gg 1$, where $\beta$ is the ratio of thermal to magnetic pressure). The observation of sharp, long-lasting temperature fluctuations in the intracluster medium (ICM) of galaxy clusters is inconsistent with classical expectations \citep{Zakamska_2003, markevitch_italchandraital_2003, richard-laferriere_constraints_2023}. A primary mechanism to generate such long-lasting fluctuations, the Field instability \citep{1965ApJ_field}, cannot produce them in the plasma outside cluster cores if the classical heat flux is present; but in the absence of thermal conduction the instability produces such fluctuations \citep[see][]{Choudhury2019MultiphaseFluctuations}. There are also significant differences between simulation predictions and experimental results in inertial confinement fusion (ICF) research, as has been outlined in \citet{rinderknecht_kinetic_2018}, and laser-plasma experiments \citep{meinecke_strong_2022}. Explaining these discrepancies is an outstanding research problem.

One of the most promising explanations is the presence of kinetic instabilities, which can play a significant role in modifying the transport properties of a plasma. Of the many kinetic instabilities characterised so far \citep[see][]{bott_kinetic_2024}, the whistler heat-flux instability (WHFI) has been identified as one of the most important because of its ability to reduce heat transport \citep{levinson_inhibition_1992, pistinner_self-inhibiting_1998, Gary_2000}. Analytical, computational, and experimental studies into the WHFI have been carried out in recent years in high-$\beta$ plasmas. These prior studies have primarily focused on moderately-high-$\beta$ plasmas that satisfy $\beta_{e} \ll L_\mathrm{T}/\rho_e$, a regime particularly relevant to astrophysical plasmas with large separations between characteristic macroscopic and microscopic length scales. In moderately-high-$\beta$ plasmas, the WHFI saturates with $\delta B^2 / B_0^2 \sim \beta_{e} \rho_e / L_\mathrm{T} \ll 1$, where $\delta B$ is the amplitude of the fluctuation in the magnetic field due to the whistler wave and $B_0$ is the magnitude of the background magnetic field, and suppression of the parallel heat flux to approximately ${\sim}$$ \beta_e^{-1}\ll 1$ times the free-streaming heat flux \citep{roberg-clark_suppression_2016, roberg-clark2018, Komarov_Schekochihin_Churazov_Spitkovsky_2018, drake_whistler-regulated_2021, meinecke_strong_2022, yerger_collisionless_2025, lopez_collisional_2025}. This implies a weakening of the power-law dependence of the heat flux on the electron temperature, reducing it from $T_e^{7/2}$ to $T_e^{1/2}$ in collisional plasmas ($\lambda_e \ll L_{\rm T}$) and from $T_e^{3/2}$ (the free-streaming scaling) to $T_e^{1/2}$ in collisionless plasmas. Recently, it has been demonstrated that such weakening also opens the possibility of generating temperature fluctuations across the ICM \citep{2026arXiv_choudhury}.

The fact that the WHFI saturates with small-amplitude fluctuations in the moderately-high-$\beta$ regime implies the existence of an ultra-high-$\beta$ regime in which $\beta_{e} \gtrsim L_\mathrm{T}/\rho_e$, and therefore $\delta B^2 / B_0^2 \gtrsim 1$. Some research has been conducted into pressure-anisotropy-driven kinetic instabilities in ultra-high-$\beta$ plasmas, such as the firehose and mirror instabilities \citep[see][]{melville_pressure-anisotropy-driven_2016, bott_thermodynamics_2025}. These studies have shown that very different physics becomes relevant in ultra-high-$\beta$ plasmas with large-amplitude fluctuations, and hence the scaling relationships discovered for the moderately-high-$\beta$ regime cannot be assumed to remain valid. Furthermore, an experiment with laser generated plasma has recently been carried out to investigate heat-flux suppression due to the WHFI in plasmas approaching the ultra-high-$\beta$ regime \citep{vincent_design_2026}; however, a theoretical description of the WHFI in this parameter space does not yet exist. The fact that this is an active area of research necessitates an accurate, analytical model for heat transport in WHFI-unstable, ultra-high-$\beta$ plasmas.

Such a model would be relevant to a range of astrophysical plasmas, including the reionised intergalactic medium of the early Universe, due to the very high values of $\beta_e$ expected to be present there \citep{barkana_beginning_2001}. There has been much interest in seed magnetic fields in the early Universe and how they became amplified to the levels seen in and around galaxies today \citep{st-onge_fluctuation_2018, st-onge_fluctuation_2020, achikanath_chirakkara_efficient_2021, zhou_spontaneous_2022, zhou_magnetogenesis_2024, hanebring_weibel_2026}. These seed fields could be primordial or Biermann-battery generated \citep{garaldi_magnetogenesis_2021} and it has been shown that kinetic instabilities could play an intermediary role in amplifying the seed magnetic field before dynamo action amplifies it further \citep{schoeffler_generation_2016}. The structure of the initial field before dynamo amplification has been shown to lead to measurable differences in present-day observations of field configurations \citep{vazza_simulations_2017, mtchedlidze_evolution_2022}, raising the possibility that such instabilities could have left identifiable markers. These magnetic fields have strong implications for galaxy formation \citep{brandenburg_galactic_2023}, hence understanding how and to what extent the WHFI may have acted on them is important.

The combination of high $\beta$ and smaller scale separation ($L_\mathrm{T}/\rho_e \sim 10^1$-$10^{3}$) in high-energy-density (HED) plasmas means they can also lie in the ultra-high-$\beta$ regime. This is of particular note to ICF research, as the large temperature gradients and weakly collisional plasmas created in ICF implosions create conditions in which the WHFI can be expected to develop, providing a mechanism for reducing energy loss from the implosion hot-spot and leading to increased hot-spot temperatures and neutron yields \citep{walsh_magnetized_2022, lopez_collisional_2025}. Magnetic fields have been proposed as a means of suppressing cross-field heat transport through magnetic confinement of electrons \citep{chang_magnetised_icf_experiment, walsh_magnetized_2022, walsh_magnetized_2024}; however, optimising an implosion for the presence of kinetic instabilities such as the WHFI also provides a mechanism for suppressing parallel transport.

It is clear, therefore, that the WHFI in ultra-high-$\beta$ plasmas requires deeper study. In this paper, we carry out particle-in-cell (PIC) simulations with OSIRIS \citep{goos_osiris_2002}, a massively parallel open-source PIC code, to answer several key questions about the WHFI and WHFI-regulated heat flux in such ultra-high-$\beta$ plasmas. We present a brief theoretical background to the WHFI in \S \ref{sec:theory}, followed by the derivation from kinetic theory of how the instability and its associated heat flux may behave in the ultra-high-$\beta$, large-amplitude regime if heat transport is dominated by diffusive (\S \ref{sec:large_whistler_diffusive}; see also \citet{ryutov_similarity_1999}) or advective (\S \ref{sec:large_whistler_advective}) processes. Our simulations of the WHFI are described in \S \ref{sec:simulations}. We perform collisionless 2D3V (two spatial dimensions and three velocity dimensions) simulations to probe the transition from the $\beta_e<L_\mathrm{T}/\rho_e$ regime to the $\beta_e > L_\mathrm{T}/\rho_e$ regime in \S \ref{sec:2d_results}. This is achieved by varying both $\beta_{e}$ and $\rho_e / L_\mathrm{T}$ independently. These simulations characterise heat flux non-locality and test the predictions made in \S \ref{sec:theory} for how magnetic-field fluctuations and the parallel heat flux saturate, as well as investigating cross-field transport. In \S \ref{sec:1d_results} we report 1D3V simulations carried out to confirm the 2D3V findings and to test whether parallel whistler waves are capable of regulating heat flux for sufficiently large values of $\delta B^2 / B_0^2$. We discuss our results and main conclusions in \S \ref{sec:discussion}.

\section{Theory of the WHFI}\label{sec:theory}
\subsection{Linear theory}\label{sec:linear_theory}

In high-$\beta$, magnetised ($\lambda_e \gg \rho_e$) plasmas, whistler waves propagating down a temperature gradient that is co-aligned with the macroscopic magnetic field can grow through cyclotron-resonant interactions with electrons if the parallel heat flux exceeds a critical value: $q_{e\|}\gtrsim q_{\rm fs}/\beta_e$~\citep{Komarov_Schekochihin_Churazov_Spitkovsky_2018,bott_kinetic_2024}. Physically, the instability arises in this scenario because the
free energy associated with the temperature gradient becomes large enough to overcome cyclotron damping. In a weakly collisional plasma, this condition on the heat flux is equivalent to the lower bound on the plasma $\beta$: that is, $\beta_e \gtrsim L_\mathrm{T}/\lambda_e$ \citep{drake_whistler-regulated_2021}. If active, the fastest-growing WHFI modes are right-hand circularly polarised and oriented parallel to the temperature gradient~\citep{bott_kinetic_2024}. If $\lambda_e/L_\mathrm{T} \gtrsim \beta_e^{-1}$, these modes have a characteristic parallel wavenumber $k_{\|} \sim \rho_e^{-1}$, and grow at a rate $\gamma_{\rm WW} \sim \Omega_e/\beta_e$, where $\Omega_e$ is the electron Larmor frequency. They also have a real frequency, with an associated phase velocity $v_{\rm ph} \sim v_{\mathrm{th}e}/\beta_e$ directed down the temperature gradient. Oblique whistler modes are also destabilised when $\beta_e \gtrsim L_\mathrm{T}/\lambda_e$, but their linear growth rate is smaller than the parallel modes by an order-unity factor. The linear theory of the WHFI has been discussed in detail elsewhere~\citep[see, e.g.,][]{bott_kinetic_2024}, so we do not discuss it further here.      

\subsection{Moderately-high-$\beta$, small-amplitude whistler regime}\label{sec:small_whistler}

After several linear growth times, previous research has shown that, if $\beta_e \ll L_\mathrm{T}/\rho_e$ -- a regime we refer to as moderately-high-$\beta$ -- the WHFI saturates while whistler waves still have a small amplitude compared to the macroscopic magnetic field: $\delta B^2 / B_0^2 \sim \beta_e \rho_e/L_\mathrm{T} \ll 1$~\citep{roberg-clark2018, Komarov_Schekochihin_Churazov_Spitkovsky_2018, yerger_collisionless_2025}. 
These studies of the WHFI in moderately-high-$\beta$ plasmas have proposed that the primary mechanism by which small-amplitude whistler waves can reduce thermal transport in a plasma is via resonant interactions with heat-carrying electrons, leading to pitch-angle scattering \citep{pistinner_self-inhibiting_1998}. Quasilinear theory predicts that the scattering frequency of resonant electrons due to these interactions with small-amplitude whistlers scales as $\nu_{\mathrm{w}} \sim \Omega_e \delta B^2 / B_0^2 \sim \beta_e v_{\mathrm{th}e}/L_{\rm T}$ \citep{yerger_collisionless_2025}. Scattering at this rate suppresses the anisotropy of the electron distribution function until the WHFI becomes marginally stable. Because whistler waves are right-hand polarised, the fastest-growing parallel whistler modes only resonate with, and therefore scatter, electrons moving up the temperature gradient, and as such are inefficient at reducing thermal conduction \citep{levinson_inhibition_1992}. Instead, oblique whistler modes are needed to scatter electrons because they have a left-hand polarisation component which can pitch-angle scatter co-propagating electrons \citep{roberg-clark_suppression_2016}. These oblique modes, in fact, become the fastest-growing due to the marginal anisotropy generated from the scattering of anti-parallel electrons \citep{bott_kinetic_2024, pistinner_self-inhibiting_1998, Komarov_Schekochihin_Churazov_Spitkovsky_2018, roberg-clark2018, choudhury_modeling_2025}. 

In this regime, the parallel heat flux becomes independent of the temperature gradient in the saturated state~\citep{pistinner_self-inhibiting_1998}:
\begin{equation}\label{eq:q_parallel}
q_{e\parallel} \sim \frac{v_{\mathrm{th}e}}{\nu_{\mathrm{w}}L_{\rm T}} q_\mathrm{fs} \sim \frac{1}{\beta_e} q_\mathrm{fs} ,
\end{equation}
as has been observed in simulations by \citet{Komarov_Schekochihin_Churazov_Spitkovsky_2018, roberg-clark2018, yerger_collisionless_2025}. 
Here,
\begin{equation}\label{eq:qfs_defn}
q_\mathrm{fs} = \int_{v_\parallel>0}^\infty \mathrm{d}^3\boldsymbol{v} \ \frac{1}{2} {m_e}v_\parallel \boldsymbol{v}^2 f_{\mathrm{M}e}(\boldsymbol{v})
\end{equation}
is the electron free-streaming heat flux, $m_e$ is the electron mass, $\boldsymbol{v}$ is particle velocity, and $f_{\mathrm{M}e}$ is a Maxwellian electron distribution function. Two different explanations for why the parallel heat flux scales in this way have been proposed. \citet{roberg-clark2018} argued that the whistler waves act as mobile scattering centres that propagate at the whistler phase velocity $v_\mathrm{ph} \sim v_{\mathrm{th}e} / \beta_e$, pitch-angle scattering electrons in that frame and limiting the advection speed of thermal energy to this value, resulting in an advective model of the heat flux. In contrast, \citet{Komarov_Schekochihin_Churazov_Spitkovsky_2018} argue that the instability saturates quasilinearly to a marginal state due to pitch-angle scattering of electrons in the frame of the background flow, limiting the distribution function anisotropy and resulting in a diffusive heat-flux model. Subsequent work by \citet{yerger_collisionless_2025} showed that both advective and diffusive contributions to the heat flux were present and contributed roughly equally to the total heat flux.

Following \citet{drake_whistler-regulated_2021}, an estimate for the perpendicular heat flux can be made using the characteristic transverse scale of whistler waves, $\rho_e$, to estimate the cross-field diffusion $D_\perp \sim \rho_e^2 \nu_\mathrm{w}$. These estimates, along with $D_\parallel \sim v_{\mathrm{th}e}^2 / \nu_\mathrm{w}$ yields a perpendicular heat flux
\begin{equation}\label{eq:q_perp}
q_{e\perp} \sim \left(\frac{\nu_{\mathrm{w}}}{\Omega_e}\right)^2 q_{e\parallel} \sim \left(\frac{\rho_e}{L_\mathrm{T}}\right)^2 \beta_e q_\mathrm{fs} .
\end{equation}

An analogous estimate can be made for the diamagnetic heat flux by assuming that this process has characteristic length and time scales determined by electron Larmor motion. This leads to a diffusion coefficient $D_\wedge \sim \rho_e^2 \Omega_e$ and heat flux
\begin{equation} \label{eq:q_diam}
q_{e\wedge} \sim \frac{\rho_e^2 \Omega_e}{v_{\mathrm{th}e}^2} \nu_\mathrm{w} q_{e\parallel}\sim \frac{\nu_{\mathrm{w}}}{\Omega_e}q_{e\parallel} \sim \frac{\rho_e}{L_\mathrm{T}}q_\mathrm{fs} .
\end{equation}

\subsection{Ultra-high-$\beta$, large-amplitude whistler regime}\label{sec:large_whistler_overview}

In ultra-high-$\beta$ plasmas with $\beta_e \rho_e / L_\mathrm{T} \gtrsim 1$, the scaling $\delta B^2 / B_0^2 \sim \beta_e \rho_e/L_\mathrm{T}$ discussed in section \S \ref{sec:small_whistler} implies that the characteristic amplitude of the magnetic field fluctuations  satisfies $\delta B / B_0 \gtrsim 1$. Thus, ultra-high-$\beta$ plasmas are expected to contain large-amplitude whistler fluctuations. In this scenario, quasilinear theory is no longer valid, and so the saturation mechanism of the WHFI might be expected to be distinct in ultra-high-$\beta$ plasmas. Large-amplitude whistler fluctuations will also interact with electrons differently than their small-amplitude counterparts, potentially causing the heat flux to depart from (\ref{eq:q_parallel}). Accordingly, a new theory of both WHFI saturation and the resulting heat flux is required in ultra-high-$\beta$ plasmas. 

To construct such a theory, we begin with the equations governing the collisionless WHFI, starting with the Vlasov equation for the electron distribution function $f_e$,
\begin{equation}
    \frac{\partial {f}_e}{\partial t} + \boldsymbol{v} \cdot \nabla {f}_e - \frac{e}{m_e} \left( \boldsymbol{E}+ \frac{\boldsymbol{v} \times \boldsymbol{{B}}}{c} \right) \cdot \frac{\partial {f}_e}{\partial \boldsymbol{v}} = 0 , 
\end{equation}
together with Faraday's law and the Maxwell-Amp\`ere law from Maxwell's equations: 
\begin{subeqnarray}
 \frac{\partial \boldsymbol{B}}{\partial t} & = & -c\bnabla \times \boldsymbol{E} , \\
   \bnabla \times \boldsymbol{B} & = &  \frac{4 \pi}{c} \boldsymbol{j} + \frac{1}{c}  \frac{\partial \boldsymbol{E}}{\partial t} .
\end{subeqnarray}
Here, $t$ is time, $\boldsymbol{E}$ is the electric field, $e$ is the elementary charge, $c$ is the speed of light, and $\boldsymbol{j}$ is the current.

To simplify the analysis, we assume that the ions are much colder than electrons ($T_i \ll T_e$), and so do not affect the dynamics of either the temperature gradient or the whistlers\footnote{If $T_{i} \sim T_e$, and $\beta_e \gtrsim (m_i/m_e)^{1/2}$ then it is a simple matter to show that the electrostatic response of the ions affects $f_e$ quantitatively but does not change any scaling relations for the heat flux we derive here, while the whistler dynamics are only weakly impacted by ions due to their small characteristic wavelength compared with the ion Larmor period.}. 
Motivated by the relationship between the heat flux, which is driven by macroscopic temperature gradients, and unstable whistler waves, we then decompose $f_e$ into two parts: a `mean' distribution function $\tilde{f}_e$ that varies on the macroscopic length scale $L_{\rm T}$ over which the temperature varies, and a perturbation $\delta f_e$ to the distribution function that describes the whistler fluctuations. To compute $\tilde{f}_e$ and $\delta f_e$ from $f_e$, we define a spatial averaging operator $\langle \cdot \rangle$, taken over scales large compared to the whistler wavelength but small compared to $L_{\rm T}$, such that any mean quantity $\tilde{X}$ is unaffected by the average, while perturbed quantities $\delta X$ vanish: namely, $\langle \tilde X \rangle = \tilde X$, while $\langle \delta X \rangle = 0$. Applying this averaging operator to the Vlasov equation yields the following pair of kinetic equations for $\tilde{f}_e$ and $\delta f_e$:
\begin{subeqnarray} \label{eq:kinetic_mean}
    \frac{\partial \tilde{f}_e}{\partial t} & + & \boldsymbol{v} \cdot \nabla \tilde{f}_e - \frac{e}{m_e} \frac{\boldsymbol{v} \times \boldsymbol{\tilde{B}}}{c} \cdot \frac{\partial \tilde{f}_e}{\partial \boldsymbol{v}} = \frac{e}{m_e} \left \langle \left( \delta \boldsymbol{E} + \frac{\boldsymbol{v} \times \delta \boldsymbol{B}}{c} \right ) \cdot \frac{\partial}{\partial \boldsymbol{v}} \delta f_e \right \rangle ,
 \label{eq:kinetic_perturbed} \\
    \frac{\partial}{\partial t} \delta f_e & + & \boldsymbol{v} \cdot \nabla \delta f_e - \frac{e}{m_e} \left( \delta \boldsymbol{E} + \frac{\boldsymbol{v} \times \delta\boldsymbol{B}}{c} \right) \cdot \frac{\partial \tilde{f}_e}{\partial \boldsymbol{v}} - \frac{e}{m_e} \frac{\boldsymbol{v} \times \boldsymbol{\tilde{B}}}{c} \cdot \frac{\partial}{\partial \boldsymbol{v}} \delta f_e \nonumber \\
    && = \frac{e}{m_e} \left(\delta \boldsymbol{E} + \frac{\boldsymbol{v} \times \delta \boldsymbol{B}}{c} \right ) \cdot \frac{\partial}{\partial \boldsymbol{v}} \delta f_e - \frac{e}{m_e} \left \langle \left( \delta \boldsymbol{E} + \frac{\boldsymbol{v} \times \delta \boldsymbol{B}}{c} \right ) \cdot \frac{\partial}{\partial \boldsymbol{v}} \delta f_e \right \rangle .
\end{subeqnarray}

For large-amplitude WHFI-unstable modes in a high-$\beta$ plasma, both the evolution equation (\ref{eq:kinetic_perturbed}\textit{b}) for $\delta f_e$ and the equations governing the perturbed electromagnetic fields can be simplified. The characteristic frequency $\omega_{\rm ww}$ of the electron-Larmor-scale whistler waves that are preferentially excited by the instability in high-$\beta$ plasmas is much smaller than the electron cyclotron frequency: $\omega_{\rm ww} \sim \Omega_e/\beta_e$. A corollary is that these whistler waves are magnetically dominated, with $|\delta \boldsymbol{E}| \sim v_{\mathrm{th}e}|\delta {\boldsymbol{B}}|/(c \beta_e)$, and the whistler phase velocity $v_{\rm ph} \sim v_{\mathrm{th}e}/\beta_e$ is much smaller than the electron thermal velocity. 
Due to the cumulative effect of interactions between electrons and the large-amplitude whistler waves, it follows that the mean distribution function is composed of an isotropic distribution $\tilde{f}_{e}^{(0)}$ and an anisotropic, heat-flux-carrying component $\tilde{f}_e^{(1)}$. Extrapolating (\ref{eq:q_parallel}) to $\beta_e \sim L_{\rm T}/\rho_e$, we find that $\tilde{f}_e^{(1)} \sim (q_{e\|}/q_{\rm fs})\tilde{f}_e^{(0)} \sim \tilde{f}_{e}^{(0)}/\beta_e$ is much smaller than $\tilde{f}_{e}^{(0)}$. Finally, we order $\delta f_e \sim \tilde{f}_e^{(1)}$ on the assumption that nonlinear wave-wave coupling could, \textit{a priori}, be important if the whistler waves have a large amplitude. Bringing these considerations together implies the following ordering of parameters: 
\begin{equation}
\frac{\omega_{\rm ww}}{\Omega_e} \sim \frac{c}{v_{\mathrm{th}e}} \frac{|\delta \boldsymbol{E}|}{|\tilde{\boldsymbol{B}}|} \sim \frac{v_{\rm ph}}{v_{\mathrm{th}e}} \sim\frac{\tilde{f}_e^{(1)}}{\tilde{f}_e^{(0)}}  \sim \frac{\delta f_e}{\tilde{f}_e^{(0)}}  \sim \frac{\rho_e}{L_{\rm T}} \sim \frac{1}{\beta_e} \ll \frac{|\delta \boldsymbol{B}|}{|\tilde{\boldsymbol{B}}|} \sim  k \rho_e \sim 1 . \label{eq:ordering}
\end{equation}
Under this ordering, (\ref{eq:kinetic_perturbed}\textit{b}) becomes, at leading order, 
\begin{multline} \label{eq:kinetic_united}
  \boldsymbol{v} \cdot \nabla \delta f_e - \tilde{\Omega}_e \frac{\partial}{\partial \varphi} \delta f_e - \frac{e}{m_e} \delta \boldsymbol{E} \cdot \frac{\partial \tilde{f}_e^{(0)}}{\partial \boldsymbol{v}} - \frac{e}{m_e c}\boldsymbol{v} \times \delta \boldsymbol{B} \cdot\frac{\partial \tilde{f}_e^{(1)}}{\partial \boldsymbol{v}} \\
    = \frac{e}{m_e c} \left( \boldsymbol{v} \times \delta \boldsymbol{B} \cdot \frac{\partial}{\partial \boldsymbol{v}} \delta f_e - \left \langle \boldsymbol{v} \times \delta \boldsymbol{B} \cdot \frac{\partial}{\partial \boldsymbol{v}} \delta f_e  \right\rangle \right ),
\end{multline}
where $\tilde{\Omega}_e \equiv e |\tilde{\boldsymbol{B}}|/m_e c$ is the frequency of Larmor motions around the macroscopic field $\boldsymbol{\tilde{B}}$, and $\varphi$ is the gyro-angle with respect to $\boldsymbol{\tilde{B}}$. Meanwhile, Faraday's law for the perturbed electromagnetic field simplifies to
\begin{equation}
 \frac{\partial }{\partial t} \delta \boldsymbol{B}= -c\nabla \times \delta \boldsymbol{E} , 
\end{equation}
while the displacement current in the Maxwell-Amp\'ere law can be neglected due to the whistlers' low frequency:
\begin{equation}
\label{eq:Ampere}
   \nabla \times \delta \boldsymbol{B} =  \frac{4 \pi}{c} \delta \boldsymbol{j} \simeq -\frac{4 \pi e}{c} \int \mathrm{d}^3\boldsymbol{v} \  \boldsymbol{v} \, \delta f_e . 
\end{equation}
The perturbed electric field can be determined by taking the first moment of (\ref{eq:kinetic_united}), self-consistently applying the ordering (\ref{eq:ordering}), and rearranging the resulting expression: 
\begin{eqnarray}
\delta \boldsymbol{E} & = & -\frac{1}{e \tilde{n}_e} \bnabla \cdot \delta \mathsfbi{P}_e + \frac{1}{4 \upi e \tilde{n}_e} \left(\bnabla \times \delta \boldsymbol{B} \right) \times \tilde{\boldsymbol{B}} \nonumber \\ && + \frac{1}{4 \upi e \tilde{n}_e} \left[ \left(\bnabla \times \delta \boldsymbol{B} \right) \times \delta {\boldsymbol{B}} -\left \langle \left(\bnabla \times \delta \boldsymbol{B} \right) \times \delta {\boldsymbol{B}} \right \rangle \right] . 
\end{eqnarray}
Here, 
\begin{equation}
\label{eq:elecpresstensor}
\delta \mathsfbi{P}_e \equiv \int\mathrm{d}^3\boldsymbol{v} \  \boldsymbol{v} \boldsymbol{v} \, \delta f_e  
\end{equation}
is the perturbed electron pressure tensor; we note that we do not need to distinguish between the total particle velocity and the peculiar velocity $\boldsymbol{w}$, because under the ordering (\ref{eq:ordering}), the perturbed electron bulk-velocity (of order the whistler phase speed) is much smaller than the electron thermal velocity. 
Faraday's law for the whistler perturbations becomes
\begin{eqnarray}
 \frac{\partial }{\partial t} \delta \boldsymbol{B} & = & \frac{c}{e \tilde{n}_e}\bnabla \times \left\{ \bnabla \cdot \delta \mathsfbi{P}_e - \frac{1}{4 \upi } \left(\bnabla \times \delta \boldsymbol{B} \right) \times \tilde{\boldsymbol{B}} \right. \nonumber \\ && \left. - \frac{1}{4 \upi} \left[ \left(\bnabla \times \delta \boldsymbol{B} \right) \times \delta {\boldsymbol{B}} -\left \langle \left(\bnabla \times \delta \boldsymbol{B} \right) \times \delta {\boldsymbol{B}} \right \rangle \right] \right\}. \label{eq:Faraday_WHFI}
\end{eqnarray}
This is a closed set of equations that completely characterises the WHFI. 

The WHFI is, in essence, driven by distribution-function anisotropy. This manifests in the evolution equation for the perturbed magnetic field via the electron pressure tensor $\delta \mathsfbi{P}_e$. Equation (\ref{eq:elecpresstensor}), in turn, implies that $\delta \mathsfbi{P}_e$ is determined directly from the perturbed electron distribution function $\delta f_e$. Thus, the way in which distribution-function anisotropy impacts $\delta f_e$ ultimately determines the behaviour of the instability. Examining the evolution equation  (\ref{eq:kinetic_united}) for $\delta f_e$, each term can be interpreted physically as follows: the streaming of particles through the perturbation, the Larmor motion of electrons around the macroscopic magnetic field, the interaction with the perturbed electric field, driven by the effect of distribution-function anisotropy, and nonlinear interactions (the final two terms). The distribution-function anisotropy provides the free energy that drives the instability. For marginality to be satisfied, this driving term needs to be balanced by one of the other terms. In the moderately-high-$\beta$ regime, this balance is provided by cyclotron damping mediated by the perturbed electric field. Under the ordering (\ref{eq:ordering}), however, nonlinear interactions becomes comparable to or larger than cyclotron damping if $\delta f_e \gtrsim f_e^{(1)}$. In the ultra-high-$\beta$ regime, we therefore posit that the saturation mechanism for the large-amplitude WHFI instability is nonlinearity.

Under this assumption, we can deduce a scaling relationship between the perturbed magnetic field and the heat flux. To establish this, a relation between $\delta B$ and $\delta f_e$ can be found from Faraday's law (\ref{eq:Faraday_WHFI}) by balancing the term responsible for the instability drive with the term responsible for nonlinear damping:
\begin{equation}
\bnabla \bcdot \delta \mathsfbi{P}_e \sim \frac{1}{4 \upi} \left[ \left(\bnabla \times \delta \boldsymbol{B} \right) \times \delta {\boldsymbol{B}} -\left \langle \left(\bnabla \times \delta \boldsymbol{B} \right) \times \delta {\boldsymbol{B}} \right \rangle \right] . \label{eq:NL_balance}
\end{equation}
Estimating the magnitude of the electron pressure tensor as $\delta \mathsfbi{P}_e \sim (\delta f_e/\tilde{f}_e^{(0)}) p_e$, using (\ref{eq:elecpresstensor}), (\ref{eq:NL_balance}) can be rearranged to give
\begin{equation} \label{eq:mag_energy_universal}
    \frac{\delta B^2}{B_0^2} \sim \beta_{e0}  \frac{\delta f}{\tilde{f}_e^{(0)}} .
\end{equation}
Here, the subscript 0 denotes the initial value of $\beta_e$. This distinction matters in the large-amplitude-whistler regime because the magnetic field at saturation can differ substantially from its initial value. For example, at saturation the local plasma beta $\beta_e$ and Larmor radius $\rho_e$ (defined using the root-mean-square (RMS) magnetic field $B_\mathrm{rms}$) are not equal to their initial values. Our assumption that nonlinear interactions cause saturation of the WHFI in ultra-high-$\beta$ plasmas then implies that $\delta f_e \sim f_e^{(1)} \sim (q_{e\|}/q_{\rm fs})f_e^{(0)}$. We deduce that
\begin{equation} \label{eq:mag_energy_universal_hf}
    \frac{\delta B^2}{B_0^2} \sim \beta_{e0}  \frac{q_{e\|}}{q_{\rm fs}} .
\end{equation}

To deduce a scaling theory for both the amplitude of the whistler perturbations and the heat flux in ultra-high-$\beta$ plasmas, we need a model of the heat flux to combine with (\ref{eq:mag_energy_universal_hf}). Motivated by the observation that, in the moderately-high-$\beta$ regime, the heat flux has both advective and diffusive components, we consider both possibilities here.

\subsubsection{Diffusive heat-flux model} \label{sec:large_whistler_diffusive}

A diffusive heat flux can be constructed following \citet{ryutov_similarity_1999}. This model assumes that electrons follow a random walk with a step size equal to the length scale on which the magnetic field changes direction. For whistler fluctuations, this distance is $\rho_e$, giving a thermal diffusivity of $\chi \simeq  \rho_e v_{\mathrm{th}e}$. The heat flux is then
\begin{equation} \label{eq:diffusive_qpredict_1}
    q_e \simeq -n_e \rho_e v_{\mathrm{th}e} \nabla T_e \sim \frac{\rho_e}{L_\mathrm{T}} q_\mathrm{fs}.
\end{equation}
Combining this with (\ref{eq:mag_energy_universal_hf}), a prediction for the perturbed magnetic energy is found:
\begin{equation} 
    \frac{\delta B^2}{B_0^2} \sim \frac{\beta_{e0} \rho_e}{L_\mathrm{T}}  \sim \left( \frac{\beta_{e0} \rho_{e0}}{L_\mathrm{T}} \right)^{2/3} .
\end{equation}
Using this to evaluate $\rho_e$ in (\ref{eq:diffusive_qpredict_1}) gives
\begin{equation} \label{eq:diffusive_qpredict}
    \frac{q_{e}}{q_\mathrm{fs}} = \left(\frac{\rho_{e0}}{L_\mathrm{T}} \right)^{2/3} \beta_{e0}^{-1/3} .
\end{equation}
Assuming that the peak wavenumber is $k \rho_e \sim 1$ as in the small-amplitude case, then it follows that the characteristic length scale of the whistler fluctuations is smaller than the Larmor radius of electrons in the initial magnetic field: 
\begin{equation}
    k \rho_{e0} \sim k \rho_{e} \frac{\delta B}{B_0} \sim \left( \frac{\beta_{e0} \rho_{e0}}{L_\mathrm{T}} \right)^{1/3} \gg 1 .
\end{equation}

To ensure this theory is self-consistent, the advective component of the heat flux must be much smaller than the diffusive component. The whistler phase velocity $v_\mathrm{ph}$ limits the advective heat flux, and if we assume that the whistler dispersion relation is the same as in the small-amplitude regime, then
\begin{equation}
    \frac{v_\mathrm{ph}}{v_{\mathrm{th}e}} \sim \frac{1}{\beta_{e0}} k \rho_{e0} \sim \left( \frac{\rho_{e0}}{L_\mathrm{T}} \right)^{1/3} \beta_{e0}^{-2/3} ,
\end{equation}
giving the ratio of the diffusive to the advective heat fluxes:
\begin{equation}
    \frac{q_e / q_\mathrm{fs}}{v_\mathrm{ph} / v_{\mathrm{th}e}} \sim \left( \frac{\beta_{e0} \rho_{e0}}{L_\mathrm{T}} \right)^{1/3} \gg 1 . 
\end{equation}
Thus, we conclude that this theory is self-consistent; the rate of advection of heat by these large-amplitude whistler fluctuations is indeed small compared with the diffusion rate. It should be noted that the assumption that the whistler dispersion relation remains the same for large-amplitude whistlers may not be true, as nonlinear effects may alter their dispersion relation.

This model predicts several key differences from the small-amplitude-whistler regime. First, the suppression of heat flux should have a much weaker dependence on $\beta_{e0}$, since $q_{e \parallel} / q_\mathrm{fs} \sim \beta_{e0}^{-1}$ in the small-amplitude regime and is only $\sim \beta_{e0}^{-1/3}$ according to the Ryutov model. Instead, the temperature-gradient length-scale $L_\mathrm{T}$ now contributes significantly to the heat flux, whereas there was no length-scale dependence in the small-amplitude regime. Secondly, the characteristic scale of whistler fluctuations in the saturated state is smaller than that found in moderately-high -$\beta$ plasmas. Finally, the heat flux would also become approximately isotropic because the magnetic field is stochastic, rendering the notion of a direction parallel to the global magnetic field largely meaningless on plasma microscales.

\subsubsection{Advective heat-flux model} \label{sec:large_whistler_advective}

Alternatively, if the advective heat flux is assumed to dominate, then the behaviour differs materially from that of the diffusive model. This advective dominance may arise if the diffusion of electrons across field lines is heavily suppressed due to the large amplitude of the magnetic field fluctuations, and instead the transport of thermal energy is limited by its advection velocity. In this case, the relevant advection velocity is the whistler phase velocity $v_\mathrm{ph}$, such that
\begin{equation} \label{eq:q_predict}
    \frac{q_{e \parallel}}{q_\mathrm{fs}} \sim   \frac{v_\mathrm{ph}}{v_{\mathrm{th}e}} ,
\end{equation}
and so
\begin{equation} \label{eq:B_predict}
    \frac{\delta B^2}{B_0^2} \sim \beta_{e0} \frac{v_\mathrm{ph}}{v_{\mathrm{th}e}} .
\end{equation}
Suppose further that the phase velocity scales with some power $\alpha$ of $\beta_{e0}$, i.e., $v_\mathrm{ph} \sim \beta_{e0}^{\alpha} v_{\mathrm{th}e}$. It then follows that
\begin{equation} \label{eq:B_predict_alpha}
    \frac{\delta B^2}{B_0^2} \sim \beta_{e0}^{1+\alpha}  .
\end{equation}
In the small-amplitude whistler regime, the phase velocity scales as $v_\mathrm{ph} \sim \beta_{e0}^{-1}$; if the phase velocity of large-amplitude whistlers were also to scale with $\beta_{e0}^{-1}$ (i.e. $\alpha = -1$), then the WHFI would saturate with 
\begin{equation} \label{eq:specB_predict}
    \delta B \sim B_0,
\end{equation}
and
\begin{equation} \label{eq:specq_predict}
    q_{e \parallel} \sim \beta_{e0}^{-1} q_{\mathrm{fs}} .
\end{equation}
In this case, the characteristic wavelength of the whistler fluctuations in saturation is then simply of order the Larmor radius of electrons in the macroscopic magnetic field.  
However, we emphasize that this value of $\alpha$ is not necessarily realised, because nonlinear effects associated with large-amplitude waves may alter the whistler dispersion relation.

As with the diffusive heat-flux model, for this theory to be self-consistent, the diffusive heat flux must be significantly smaller than the advective component:
\begin{equation}
    \frac{v_{\mathrm{th}e}}{\nu_\mathrm{eff} L_\mathrm{T}} \ll \frac{v_\mathrm{ph}}{v_{\mathrm{th}e}} . \label{eq:diffweakasump}
\end{equation}
For $\alpha = -1$, this requires
\begin{equation}
     \frac{\nu_\mathrm{eff}}{ \Omega_{e0}} \gg \frac{\beta_{e0}\rho_{e0}}{L_\mathrm{T}} \gtrsim 1 .
\end{equation}
Thus, the scattering of electrons by large-amplitude whistlers must strongly disrupt Larmor motion caused by the macroscopic magnetic field in some manner. 

The advective heat-flux model predicts behaviour that differs substantially from the diffusive heat-flux model and is instead much closer to that found in moderately-high-$\beta$ plasmas. Unlike the diffusive model, the advective model predicts that neither the heat flux nor the saturated magnetic field strength depends on the temperature-gradient length-scale. However, the extent to which heat flux is suppressed depends on how the phase velocity scales with $\beta_{e0}$ for large-amplitude waves; if $v_{\rm ph}$ is still inversely proportional to $\beta_{e0}$, then the advective heat flux retains the scaling (\ref{eq:q_parallel}).  
Another important distinction between the two models is that, in the advective model, cross-field transport continues to follow the Braginskii scalings~\citep{braginskii_transport_1965}, whereas the diffusive model predicts an approximately isotropic heat flux. 

The simulations presented in the following sections test these predictions and determine which of the two transport models is realised in the ultra-high-$\beta$ regime.

\section{Simulations}\label{sec:simulations}
\subsection{Numerical setup}
\begin{table}
    \centering
    \begin{tabular}{|c|c|c|c|c|c|c|c|c|}
    \cline{1-9}
    Dimensions & $\theta \  (^\circ )$ & $L_x / \rho_{e0}$ & $L_y / \rho_{e0}$ & $L_{\mathrm{T}0} / \rho_{e0}$  & $n_{\mathrm{ppc}}$ & $\beta_{e0}$ & $t_\mathrm{end} \Omega_{e0} (\times 10^3)$ & $t_\mathrm{end} \omega_{p0} (\times 10^3)$\\ \cline{1-9}

    \multirow{15}{*}{2} & \multirow{10}{*}{0}  & \multirow{7}{*}{50} & \multirow{10}{*}{10} & \multirow{7}{*}{100}   & \multirow{10}{*}{900}  & 20 & $12$ & 190 \\ \cline{7-9}
     &  &  &  &  &  & 40 & $4.9$ & 110  \\ \cline{7-9}
     &  &  &  &  &  & 60 & $1.6$ & 43  \\ \cline{7-9}
     &  &  &  &  &  & 80 & $1.4$ & 43  \\ \cline{7-9}
     &  &  &  &  &  & 120 & $3.0$ & 120 \\ \cline{7-9}
     &  &  &  &  &  & 180 & $2.8$ & 130 \\ \cline{7-9}
     &  &  &  &  &  & 280 & $3.5$ & 200 \\ \cline{3-3}  \cline{5-5}\cline{7-9}
     &  & 22 &  & 44  &  &  \multirow{3}{*}{120} & $2.4$ & 92 \\ \cline{3-3} \cline{5-5} \cline{8-9}
     &  & 33 &  & 66  &  &  & $1.6$ & 63 \\ \cline{3-3}  \cline{5-5} \cline{8-9}
     &  & 75 &  & 150 &  &  & $3.1$ & 120 \\ \cline{2-9}
     & \multirow{5}{*}{45}  & \multirow{5}{*}{35} &  \multirow{5}{*}{7.1} &  \multirow{5}{*}{100} & \multirow{4}{*}{900} & 40 & $5.8$ & 130 \\ \cline{7-9}
     &  &  &  &  &  & 120 & $2.8$ & 110 \\ \cline{7-9}
     &  &  &  &  &  & 180 & $3.7$ & 180 \\ \cline{7-9}
     &  &  &  &  &  & 280 & $2.7$ & 160 \\ \cline{6-9}
     &  &  &  &  & 784  & 400 & $2.7$ & 190 \\ \cline{1-9}     
    
    \multirow{15}{*}{1}  & \multirow{15}{*}{0}  & \multirow{10}{*}{50} & \multirow{15}{*}{-} & \multirow{10}{*}{100}   & \multirow{15}{*}{3600}   & 20 & $8.9$ & 140  \\ \cline{7-9} 
     &  &  &  &  &  & 40 & $5.1$ & 110  \\ \cline{7-9} 
     &  &  &  &  &  & 60 & $7.2$ & 200  \\  \cline{7-9}
     &  &  &  &  &  & 80 & $5.6$ & 180  \\ \cline{7-9}
     &  &  &  &  &  & 120 & $3.7$ & 140 \\ \cline{7-9}
     &  &  &  &  &  & 180 & $4.4$ & 210 \\ \cline{7-9}
     &  &  &  &  &  & 280 & $3.1$ & 180 \\ \cline{7-9}
     &  &  &  &  &  & 400 & $4.7$ & 330 \\ \cline{7-9}
     &  &  &  &  &  & 600 & $5.7$ & 500 \\ \cline{7-9}
     &  &  &  &  &  & 1000 & $6.6$ & 740 \\ \cline{3-3} \cline{5-5} \cline{7-9}
     &  & 22  &  & 44 & & \multirow{5}{*}{180} & $10$ & 480 \\ \cline{3-3} \cline{5-5} \cline{8-9}
     &  & 33  &  & 66 &  &  & $7.0$ & 330 \\ \cline{3-3} \cline{5-5} \cline{8-9}
     &  & 75  &  & 150 &  &  & $4.6$ & 220 \\ \cline{3-3} \cline{5-5} \cline{8-9}
     &  & 110 &  & 225 &  &  & $7.4$ & 350 \\ \cline{3-3} \cline{5-5} \cline{8-9}
     &  & 170  &  & 340 &  &  & $13$ & 610 \\ \cline{1-9}
    \end{tabular}
    \caption{Initial parameters for the simulations performed in this study. $\theta$ is the angle between the temperature gradient and the applied magnetic field, $n_\mathrm{ppc}$ is the number of particles per cell, $t_\mathrm{end}$ is the duration of the simulation, and $\omega_{\mathrm{p}0}$ is the plasma frequency at the hot boundary of the simulation.}
    \label{tab:simulations}
\end{table}

 All simulations were initialised with isobaric initial conditions similar to those used in \citet{yerger_collisionless_2025}, with 
\begin{equation} \label{eq:ICs}
 n(x) = n_{0} \left(\frac{2}{1 + x /L_x} \right), \qquad
 T(x) = T_\mathrm{c}\left(1+\frac{x}{L_x}\right) . 
\end{equation}
Here, $n$ is the number density of electrons, $T$ is the electron temperature, $T_c$ is the electron temperature at the cold (left-hand) boundary, and $L_x$ is the length of the simulation box in the direction parallel to the temperature gradient (the $x$ direction). Throughout this paper, the subscript $0$ denotes an initial value at the hot (right-hand) boundary. The reference free-streaming heat flux is taken to be its value at the hot boundary of the simulations: $q_\mathrm{fs} = 0.030 n_e m_e c^3$. The electron temperature and density are equal to the ion temperature and density at all locations.

For all simulations, thermal-bath boundary conditions were used at the left and right walls, such that any particle leaving the simulation domain at that boundary is immediately reinjected at the same location with a thermal velocity randomly sampled from a half-Maxwellian distribution for the velocity component perpendicular to the wall, and Maxwellian distributions for the other velocity components. This allowed us to maintain boundary temperatures of $T_\mathrm{c} = 10.2\mathrm{keV}$ and $T_\mathrm{h} = 20.4\mathrm{keV}$, respectively, such that $\sqrt{T_\mathrm{h} / m_e} =0.2$. The top and bottom boundary conditions were periodic. A realistic proton-to-electron mass of $1836 m_e$ was used to simulate a proton-electron plasma and the grid spacing $\Delta x$ was chosen such that $\lambda_{\rm D} = 1.5 \Delta x$ to minimise spurious particle heating. All other simulation parameters are listed in table \ref{tab:simulations}. The number of particles per cell $n_{\mathrm{ppc}}$ was chosen to ensure that numerical noise was negligible, giving $n_{\mathrm{ppc}} =3600$ in 1D3V and $n_{\mathrm{ppc}}=900$ in 2D3V. The only exception to this was for the 2D3V simulation with $\beta_{e0}=400$, where a smaller $n_{\mathrm{ppc}}$ was chosen to make the simulation run time feasible. $L_x = L_{\mathrm{T}0}/2$ and $L_y = L_{\mathrm{T}0}/10$ (except for simulations with varying $L_{\mathrm{T}0} / \rho_{e0}$, for which $L_y = 10 \rho_{e0}$). For simulations with the magnetic field inclined by an angle $\theta$, both box dimensions were additionally divided by $\cos{\theta}$.

For the electromagnetic-field boundary conditions, we implemented a masking function in OSIRIS that decreases the field strengths near the thermal bath boundaries following \citet{umeda_improved_2001}, as subsequently adopted by \citet{Komarov_Schekochihin_Churazov_Spitkovsky_2018} and \citet{ yerger_collisionless_2025}. The masking function is of the form
\begin{equation}\label{eq:masking}
    f_{\mathrm{Mask}}(x, L_D, r) = 
        \begin{cases}
            1 - \left(r\frac{x-L_D}{L_D}\right)^2, &  x \leq L_D,\\
            1, &  L_D < x < L_x - L_D,\\
            1 - \left(r\frac{x-L_x+L_D}{L_D}\right)^2, & x \geq L_x-L_D,
        \end{cases}
\end{equation}
where $r$ controls the strength of the damping and $L_D$ is the length of the masking region. The values $L_D = 2\rho_{e0}$ and $r=0.25$ were chosen to minimise the fraction of whistler waves reflected at the boundary. 

The ratio $L_{\mathrm{T}0} / \rho_{e0}$ was varied by changing $L_x / \rho_{e0}$ while keeping the temperature at each boundary the same. The initial plasma beta $\beta_{e0} = 8\pi n_0 T_\mathrm{h} / B_0^2$ was varied by applying a uniform background magnetic field across the simulation domain. Decreasing $B_0$ increases $\beta_{e0}$; however, it also increases $\rho_{e0}$, hence larger simulation domains are required for higher $\beta_{e0}$ simulations, making them more computationally expensive. Because the simulations were initialised with an isobaric equilibrium, $\beta_{e0}$ is constant throughout the simulation domain. Each simulation was run until the WHFI reached saturation, which also takes longer for high $\beta_{e0}$ simulations. These increased computational demands limit the maximum value of $\beta_{e0}$ that can be investigated. 

The simulations were successfully benchmarked against the TRISTAN-MP simulations of \citet{yerger_collisionless_2025} by comparing the time evolution of the magnetic-field strength and parallel heat flux for simulations with similar initial parameters.

\subsection{2D3V results}\label{sec:2d_results}
\subsubsection{Varying plasma beta}\label{sec:2d_beta_results}

\begin{figure}
\centering
\includegraphics[width=\textwidth]{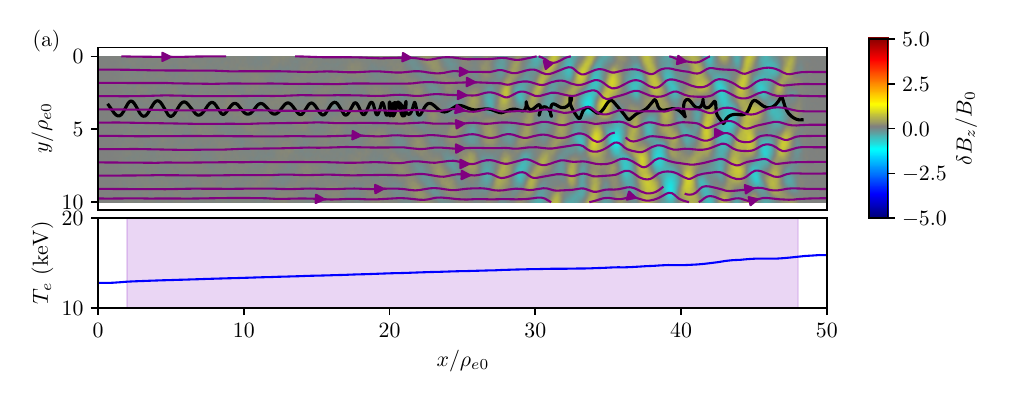}
\includegraphics[width=\textwidth]{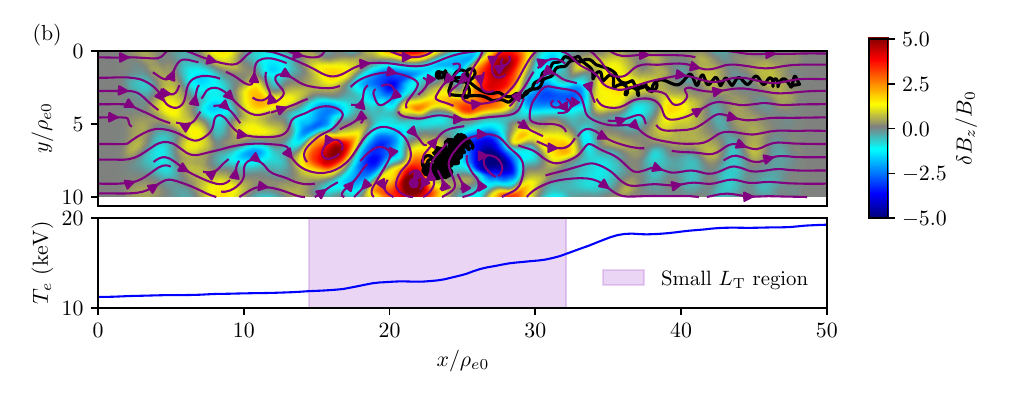}
\caption{The magnetic fields and $T_e$ from a 2D3V WHFI simulation with $\beta_{e0} = 180$ during (a) exponential growth at $t\Omega_{e0} = 60$ and (b) nonlinear growth at $t\Omega_{e0} = 1069$. The temperature increases along the $x$ direction, and the background magnetic field is parallel to the $x$ axis. Panel (a) shows the formation of waves with wavefronts oblique to the background magnetic field in the hotter (right-hand) half of the simulation box and how they propagate to the colder half in b). Black lines show trajectories of test electrons traced around the times corresponding to the background-field images, demonstrating magnetic trapping and reflections when they interact with whistler waves. Purple shaded regions show the domain of the small-$L_\mathrm{T}$ region at both times.
}
\label{fig:beta180_waves}
\end{figure}

\begin{figure}
\centering
\includegraphics[width=0.7\textwidth]{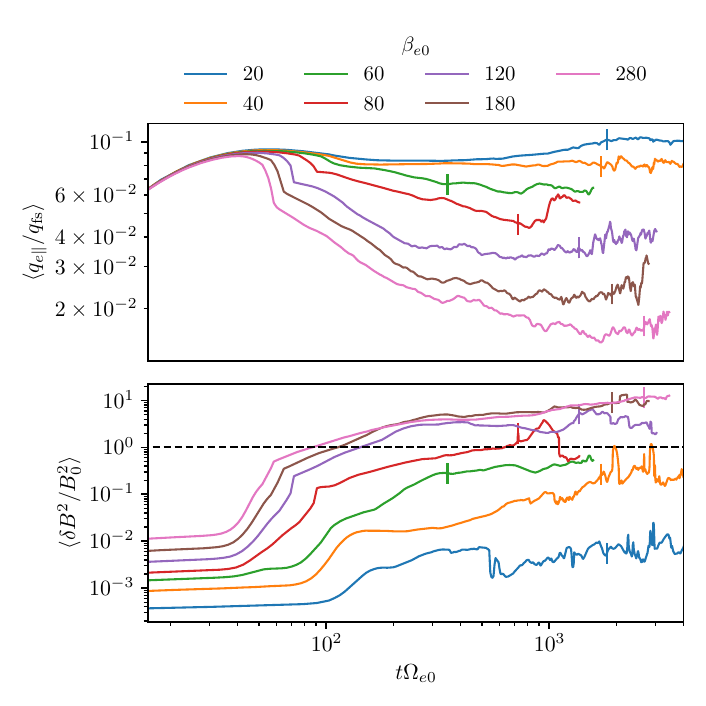}
\caption{Temporal evolution of the spatially averaged parallel heat flux (upper) and perturbed magnetic energy (lower) for 2D3V simulations with background magnetic fields parallel to the temperature gradient, $L_{\mathrm{T}0} / \rho_{e0} = 100$, and a range of $\beta_{e0}$ values. Both quantities are spatially averaged over the small-$L_\mathrm{T}$ region and the heat flux is normalised to the free-streaming value. A black line is drawn at $\langle \delta B^2 / B_0^2 \rangle= 1$ to show when the perturbed magnetic energy exceeds that of the background field. Markers indicate the time at which saturation was determined to be have been reached (the start of the saturated interval for each simulation).}
\label{fig:2d_time_evol}
\end{figure}

We first performed 2D3V simulations of the WHFI designed to access the ultra-high-$\beta$ regime by increasing $\beta_{e0}$ at fixed $L_{\mathrm{T}0} / \rho_{e0}$. Values of $\beta_{e0} \in \{20, 40, 60, 80, 120, 180, 280\}$ were used, with $L_{\mathrm{T}0} / \rho_{e0}=100$ and an external magnetic field applied parallel to the temperature gradient in the $x$ direction. The magnetic field for a sample simulation ($\beta_{e0}=180$) is shown in figure \ref{fig:beta180_waves}, demonstrating the growth and saturation of the WHFI in the ultra-high-$\beta$ regime. The waves initially form in the hotter half of the simulation before propagating to the colder half. Superimposing trajectories of a sample of electrons on these field snapshots shows that some electrons become trapped in low-field regions of the plasma. This demonstrates that, in addition to resonant scattering, large-amplitude waves possess an alternative mechanism for suppressing the heat flux: variations in magnetic-field strength are large enough to form effective magnetic mirrors. The large perturbations are also capable of forming locally closed or strongly distorted magnetic field lines that trap electrons. 

The region in the simulation in which the magnetic energy is maximised is found to correspond strongly with the location of steepest temperature gradient, consistent with the formation of a transport barrier by the whistler waves and confirming that the fluctuations draw energy from the temperature gradient (see figure \ref{fig:beta180_waves}). A `small-$L_\mathrm{T}$ region' was defined at each time step as follows. We averaged the temperature over the $y$ direction and smoothed the result, which gives the temperature profile shown in the bottom panels of figure \ref{fig:beta180_waves}. We then calculated the local $\rho_{e0} / L_\mathrm{T}$ at each $x$ coordinate using the gradient of this temperature profile, which yielded a peaked profile once large enough whistlers are present. Taking the full width at half maximum (FWHM) of a Gaussian fit to this $\rho_{e0} / L_\mathrm{T}$ profile then yielded the domain of the small-$L_\mathrm{T}$ region. The small-$L_\mathrm{T}$ region was constrained not to include the masking regions near the boundaries, and was found to propagate down the temperature gradient with the whistler waves. This defines the region of interest for exploring the WHFI, as it is here that the temperature gradient is steep enough to strongly destabilise whistler waves. The temporal evolution of the heat flux (normalised to the free-streaming value, defined for the particle distribution at the hot wall of the simulation) and perturbed magnetic energy are therefore both averaged over this small-$L_\mathrm{T}$ region. The resulting time histories are shown in figure \ref{fig:2d_time_evol}.

Because the simulation is initialised without any heat flux, a time interval of order ${\sim}L_x/v_{\mathrm{th}e}$ is required before the heat flux can develop. Once this has happened, the WHFI is destabilised and starts to grow, causing a corresponding decrease in $\langle q_{e\parallel} \rangle$, where angle brackets here denote a spatial average over the small-$L_\mathrm{T}$ region. The initial decrease in the heat flux once the whistler waves are destabilised appears very sharp partly because this is when the small-$L_\mathrm{T}$ region becomes well-defined and is no longer the entire spatial domain. Thus, the rapid change reflects both a physical decrease in the heat flux and a rapid change in the spatial region over which the heat flux is  averaged. A similar effect affects the preliminary growth of the magnetic fields. Initially the wave growth is exponential; however, after some time, this growth slows down and continues at a greatly decreased rate until saturation is reached. Figure \ref{fig:2d_time_evol} shows that $\delta B^2 / B_0^2 > 1$ is reached, indicating that the WHFI-perturbed magnetic field becomes comparable to, or larger than, the background field, but only by an order-unity factor.

\begin{figure}
    \centering
    \includegraphics[width=0.49\linewidth]{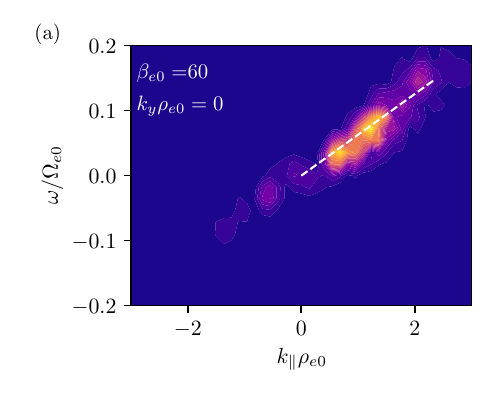}
    \includegraphics[width=0.49\linewidth]{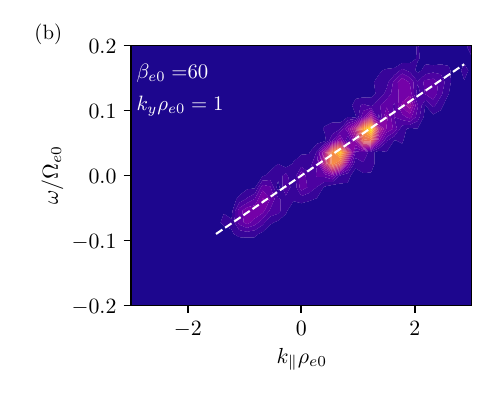}
    \includegraphics[width=0.49\linewidth]{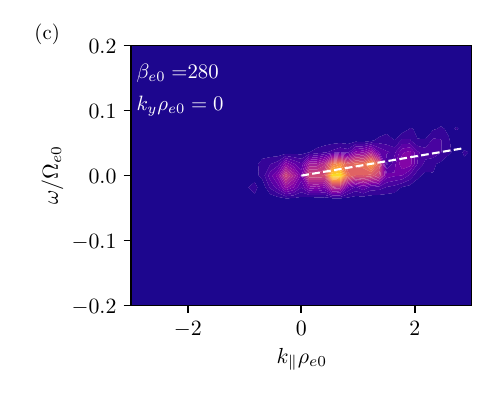}
    \includegraphics[width=0.49\linewidth]{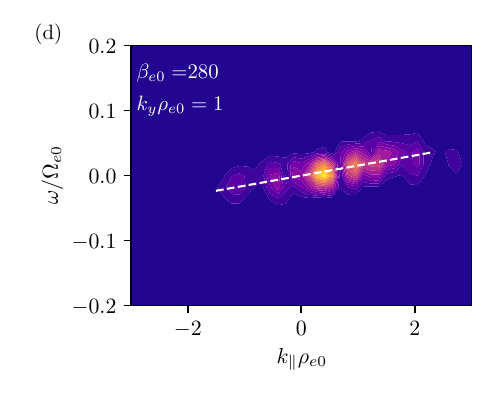}
    \caption{Spectrograms in wavenumber-frequency space of $B_y + \mathrm{i}B_z$ for panels (a),(b) $\beta_{e0} = 60$ and (c),(d) $\beta_{e0} = 280$ in the saturated state. The plots are taken from magnetic field data over the entire spatial domain, excluding the boundary masking regions, and over the last $350 \Omega_{e0}^{-1}$ of simulation time. Plots for both parallel ($k_y \rho_{e0}=0$) and oblique ($k_y \rho_{e0}=1$) wavenumbers are shown in (a),(c) and (b),(d), respectively. The linear best fit for the pixels with maximum intensity at each wavenumber is also shown in each panel, following $\omega / \Omega_{e0} = \hat{v}_\mathrm{ph} k_\parallel \rho_{e0}$. For parallel modes, the fits are restricted to $k_\parallel \rho_{e0} > 0$ since parallel modes only resonantly interact with electrons moving anti-parallel to the background magnetic field, leading to much higher power in this region.}
    \label{fig:2dbeta_dispersion_plots}
\end{figure}

To characterise the phase velocity of the whistler waves, we calculate the dispersion relation for the different values of $\beta_{e0}$ investigated. This is done by carrying out a two-dimensional Fourier transform of the magnetic field fluctuation $B_y + \mathrm{i}B_z$ at every point in time. The mode of interest is than extracted (either $k_y \rho_{e0} = 0$ for parallel modes or $k_y \rho_{e0} = 1$ for oblique modes) and then a Fourier transform in time is carried out on this mode, to produce the spectrograms in figure \ref{fig:2dbeta_dispersion_plots}. We find that, in the saturated state, the dispersion relation in this large-amplitude regime differs from that measured from simulations in the small-amplitude regime~\citep{yerger_collisionless_2025}. To demonstrate this, the dispersion relations for representative simulations ($\beta_{e0} \in \{60, 280\}$) are plotted in figure \ref{fig:2dbeta_dispersion_plots}. These show that the dispersion relation is linear, following $\omega / \Omega_{e0} = \hat{v}_\mathrm{ph} k_\parallel \rho_{e0}$, where $\hat{v}_\mathrm{ph} = v_\mathrm{ph} / v_{\mathrm{th}e}$ is a fitting parameter which is the phase velocity normalised by the electron thermal velocity. The substantial decrease in slope between $\beta_{e0}=60$ and $\beta_{e0}=280$ shows qualitatively that $\hat{v}_\mathrm{ph}$ decreases with increasing $\beta_{e0}$ for both parallel and oblique modes. 

\begin{figure}
    \centering
    \includegraphics[width=\linewidth]{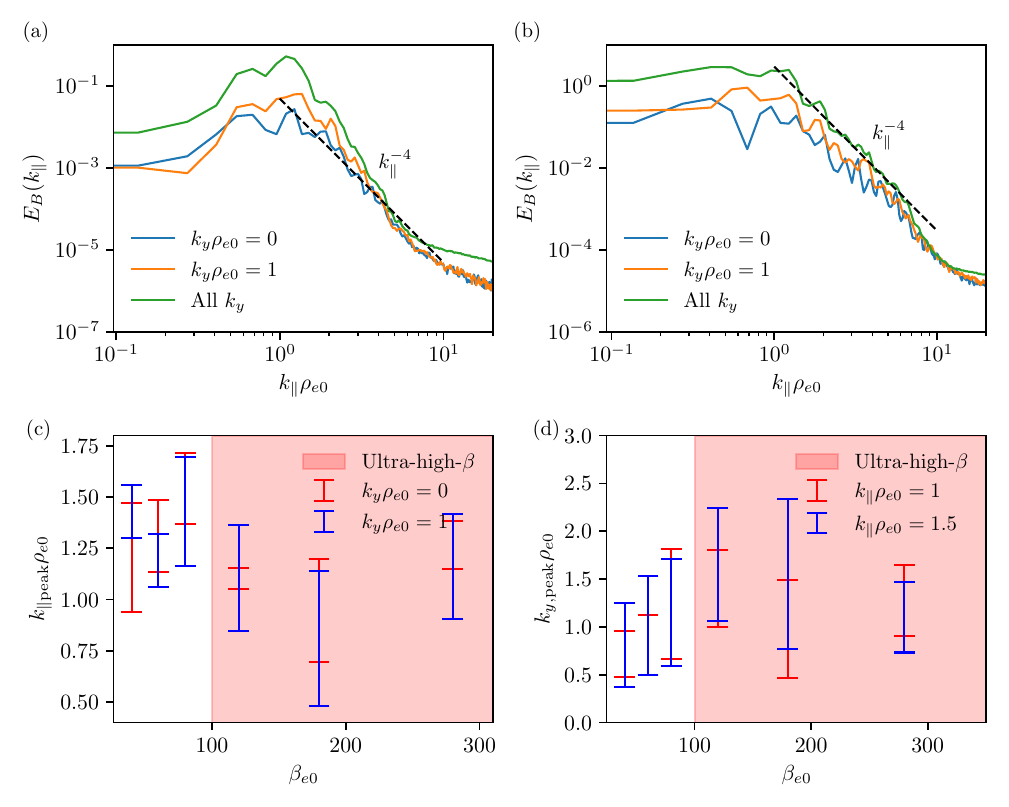}
    \caption{Fourier spectra of $B_y + \mathrm{i}B_z$ for 2D3V simulations with $L_{\mathrm{T}0} / \rho_{e0} = 100$ and a parallel macroscopic magnetic field with (a) $\beta_{e0} = 60$ and (b) $\beta_{e0} = 280$. Shown are spectra for parallel ($k_y \rho_{e0} = 0$) and oblique ($k_y \rho_{e0} = 1$) modes, as well as spectra integrated over all $k_y$. The spectra are normalised such that $\int \mathrm{d}k_\parallel \ E_B (k_\parallel) = \langle \delta B_y^2 +\delta B_z^2 \rangle / B_0^2$, the perturbed magnetic energy perpendicular to the applied field for those modes. As in figure \ref{fig:2dbeta_dispersion_plots}, the data are taken over the entire spatial domain, excluding the boundary masking regions, and the last $350 \Omega_{e0}^{-1}$ of each simulation. The $k_\parallel^{-4}$ scaling typical of whistler waves is shown over its relevant region. c) The peak parallel wavenumber for both oblique and parallel modes for all of the $\beta_{e0}$ values tested, where the peak wavenumber (the wavenumber at which $k_\parallel E_B(k_\|)$ peaks) in the last $350 \Omega_{e0}^{-1}$ of each simulation and the mean of these values is plotted. Error bars show the standard deviation of this value. The same if plotted in (d) but for the peak perpendicular wavenumber for modes with $k_\parallel \rho_{e0} \in \{1,1.5\}$}
    \label{fig:2dbeta60_kspecra}
\end{figure}

The Fourier spectra of the whistler waves are shown in figure \ref{fig:2dbeta60_kspecra} for both parallel and oblique modes for $\beta_{e0}=60$ and $\beta_{e0}=280$, which appear very similar to those observed in the small-amplitude regime. Notably, the spectra continue to demonstrate $k_\parallel^{-4}$ scaling, as has been observed by \citet{roberg-clark2018, yerger_collisionless_2025, choudhury_modeling_2025}. The Fourier spectra integrated over $k_y$ are also shown. For $\beta_{e0} = 60$, the integration over all modes reveals some unexpected phenomena: a steeper gradient to the spectrum, a more pronounced peak at $k_\parallel \rho_{e0} \sim 1$, and more power in the high-wavenumber tail. This may indicate cyclotron resonance playing a role near the transition between the small- and large-amplitude regimes.

The peak parallel wavenumber was defined as the wavenumber at which the wave energy is maximal (i.e. the wavenumber at which $k_\parallel E_B(k_\|)$ peaks). This value is also plotted in figure \ref{fig:2dbeta60_kspecra}, for all values of $\beta_{e0}$ simulated. The peak parallel wavenumber remains around $k_\parallel \rho_{e0} \sim 1$ for all values of $\beta_{e0}$. This may seem to contradict the expectation that $k_\parallel \rho_{e} \sim 1$; yet, since $\delta B / B_0$ does not exceed order unity, $k_\parallel \rho_{e} \sim k_\parallel \rho_{e0}$ and consequently the expectation remains valid. The peak perpendicular wavenumber $k_{y, \mathrm{peak}} \rho_{e0}$ was calculated in the same way and is also plotted in figure \ref{fig:2dbeta60_kspecra}, also remaining around $k_y \rho_{e0} \sim 1$.

\begin{figure}
\centering
\includegraphics[width=\textwidth]{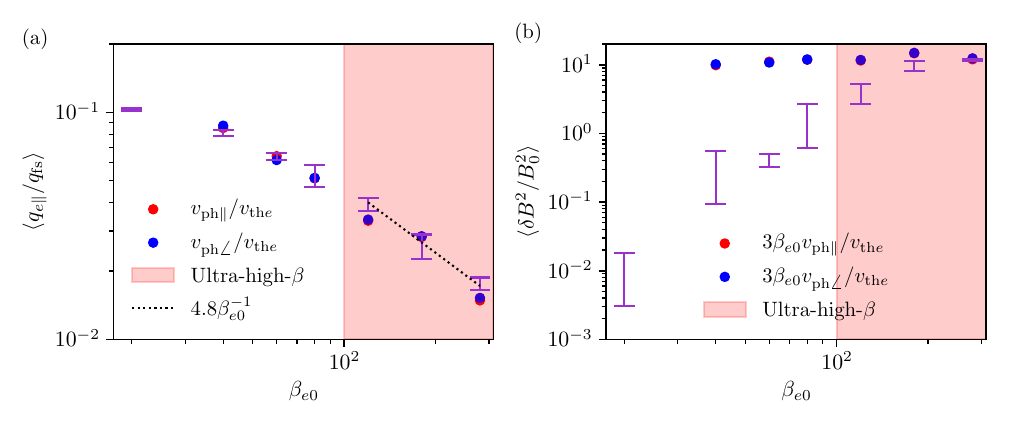}
\caption{Scalings of (a) the heat flux and (b) the perturbed magnetic energy with $\beta_{e0}$ in 2D3V simulations with $L_{\mathrm{T}0} / \rho_{e0}=100$, where the values on the $y$ axes have been spatially averaged over the small-$L_\mathrm{T}$ region. In panel (a), the best-fitting line of the form $\langle q_{e\parallel}/q_{\mathrm{fs}} \rangle = A\beta_{e0}^{-1}$, where $A \approx 4.8$ is a fitting parameter, is shown by the black dotted line in the ultra-high-$\beta$ region. The expected values calculated from the whistler phase velocities, found using the best fit to the dispersion relation at each $\beta_{e0}$, are also shown for both oblique ($v_{\mathrm{ph}\angle}$, corresponding to modes with $k_y \rho_{e0} = 1$) and parallel ($v_{\mathrm{ph}\parallel}$, corresponding to modes with $k_y \rho_{e0} = 0$) modes. The predictions in b) have been multiplied by a best-fitting order-unity prefactor. The lowest $\beta_{e0}$ simulation ($\beta_{e0}=20$) did not strongly destabilise the WHFI, and hence no dispersion relation could be found.}
\label{fig:2d_scaling}
\end{figure}

Because the large-amplitude whistler waves have an approximately linear dispersion relation, the phase velocity no longer depends on the peak wavenumber, and instead $v_\mathrm{ph} = \hat{v}_\mathrm{ph}v_{\mathrm{th}e}$. For an advective heat flux, the perturbed magnetic energy is expected to follow $\delta B^2 / B_0^2 \sim \beta_{e0} v_\mathrm{ph} /v_{\mathrm{th}e}$ and the heat flux is expected to follow $q_{e\parallel} / q_{\mathrm{fs}} \sim v_\mathrm{ph} /v_{\mathrm{th}e}$. The scalings $\langle \delta B^2 / B_0^2 \rangle$ and $\langle q_{e\parallel}/q_{\mathrm{fs}} \rangle$ with $\beta_{e0}$ are plotted in figure \ref{fig:2d_scaling}, where these quantities have been spatially averaged over the small-$L_\mathrm{T}$ region. Theoretical predictions based on the measured values of $v_\mathrm{ph} / v_{\mathrm{th}e}$ at each $\beta_{e0}$, together with (\ref{eq:q_predict}) and (\ref{eq:B_predict}), are also shown. Predictions inferred from both parallel ($v_{\mathrm{ph}\parallel}$) and oblique ($v_{\mathrm{ph}\angle}$) modes are plotted and shown to be very similar.

The theoretical predictions made in \S \ref{sec:large_whistler_advective} for the advective heat-flux model, rather than the diffusive heat-flux model outlined in \S \ref{sec:large_whistler_diffusive}, are a close match to the simulation results. In particular, $q_{e\parallel}/q_{\mathrm{fs}}$ agrees with $v_\mathrm{ph} / v_{\mathrm{th}e}$ for both oblique and parallel modes to within an order-unity prefactor. The more specific prediction (\ref{eq:specq_predict}), obtained by assuming that the large-amplitude whistler dispersion relation remains close to its small-amplitude form, also appears to hold: $q_{e\parallel}/ q_{\mathrm{fs}} \sim \beta_{e0}^{-1}$ describes the heat flux results well. Given this scaling of the heat flux and phase velocity, it would be expected that $\delta B \sim B_0$, and indeed for $\beta_{e0}\rho_{e0}/L_{\mathrm{T}0} \gtrsim 1$, the perturbed magnetic energy plateaus at a value corresponding to $\delta B \sim B_0$ ($\delta B_{\rm rms} \approx 3.3 B_0$). Because of the large computational expense of running such simulations, it was not possible to investigate larger values of $\beta_{e0}$ with the same initial conditions; however, a plateau at a similar saturation value is also observed in the angled simulations in figure \ref{fig:ang_scaling}, which investigated larger values of $\beta_{e0}$ (see \S \ref{sec:angled_results}). These results support the idea that heat transport becomes primarily advective when large-amplitude whistler waves are destabilised. We note that the power-law scaling of $\delta B^2 / B_0^2$ with $\beta_{e0}$ in the moderately-high-$\beta$ regime is steeper than the scaling found in previous studies~\citep{Komarov_Schekochihin_Churazov_Spitkovsky_2018, yerger_collisionless_2025}. We believe that this is because our simulations were not designed to characterise the moderately-high-$\beta$ regime; the range in $\beta_{e0}$ between the onset of heat-flux suppression by the WHFI and the transition to the ultra-high-$\beta$ regime is too narrow for the $\delta B^2 / B_0^2 \propto \beta_e$ scaling to be resolved. 

\begin{figure}
\centering
\includegraphics[width=0.7\textwidth]{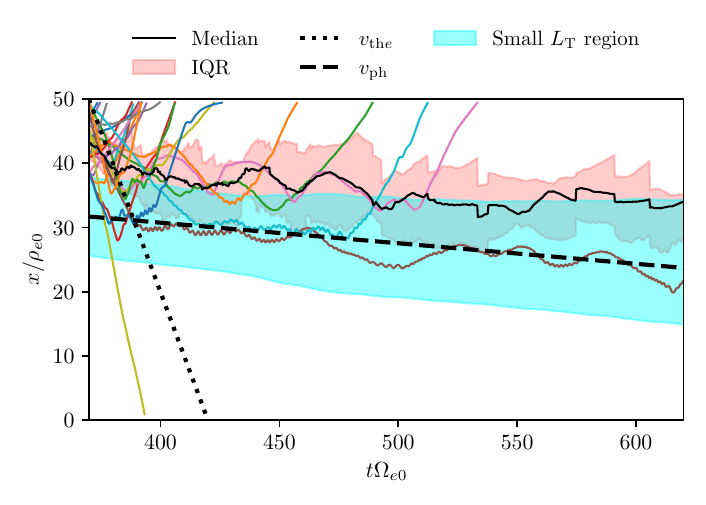}
\caption{Electron trajectories for particles that were on the hot side of the small-$L_\mathrm{T}$ region at $t\Omega_{e0}=370$ in the 2D3V simulation with $\beta_{e0}=120$ and $L_{\mathrm{T}0} / \rho_{e0} = 100$. Only a subset of electrons that met these criteria had their trajectories plotted to avoid overcrowding the figure (coloured lines). However, the plotted median and interquartile range (IQR) of electron positions are computed using all of the electrons that met the criteria. The position of the small-$L_\mathrm{T}$ region at each time is shown in blue. The whistler phase velocity $v_\mathrm{ph}$, calculated from the linear best fit to the dispersion relation for parallel modes from the same simulation, is also shown. The thermal velocity is also plotted for reference.}
\label{fig:2d_barrier}
\end{figure}

One important aspect of the advective heat-flux model proposed in \S \ref{sec:large_whistler_advective} is that electrons travelling from the hot wall to the cold wall  are unable to transport thermal energy towards the cold wall faster than $v_\mathrm{ph}$ via diffusion [cf. (\ref{eq:diffweakasump})]. Electron trajectories are shown in figure \ref{fig:2d_barrier}, showing that although a small fraction of particles  pass through the small-$L_\mathrm{T}$ region, the majority of particles are either reflected back to the hot wall or become trapped in the small-$L_\mathrm{T}$ region and are advected at approximately the phase velocity. As such, the small-$L_\mathrm{T}$ region acts as an effective transport barrier, as was suggested by figure \ref{fig:beta180_waves}. Indeed, of the particles tracked in figure \ref{fig:2d_barrier}, only 12\% made it through the transport barrier, and only 4.9\% made it through unaffected (defined as reaching the cold wall without travelling towards the hot wall at any point in their trajectory). The electrons that were unaffected by the transport barrier were near-thermal electrons with an average speed of $v / v_{\mathrm{th}e} = 1.2 \pm 0.5$ and, as such, did not carry significant heat flux. This provides strong evidence that thermal energy cannot be transported down the temperature gradient faster than the whistler phase speed in such a system.

\subsubsection{Varying temperature gradient}\label{sec:2dLTresults}

\begin{figure}
\centering
\includegraphics[width=0.7\textwidth]{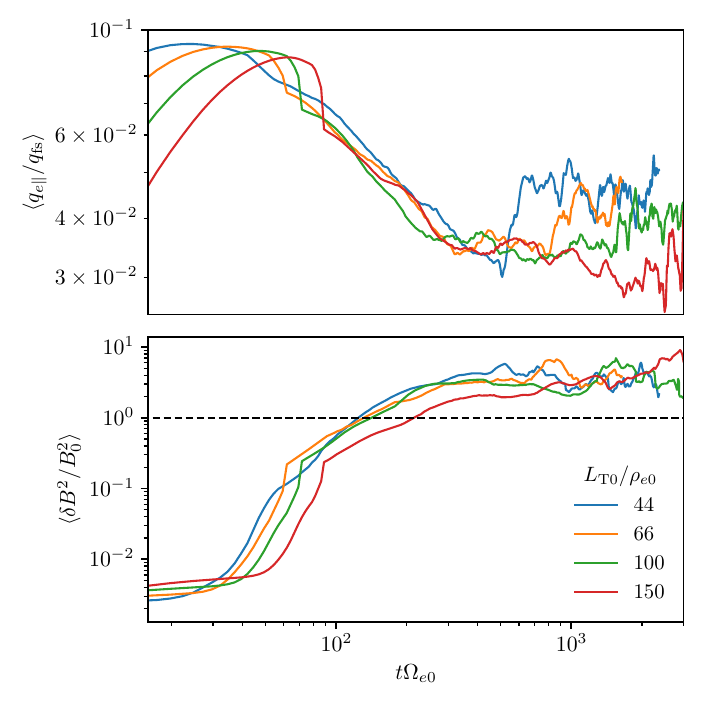}
\caption{Temporal evolution of the spatially averaged parallel heat flux (upper) and perturbed magnetic energy (lower) for 2D3V simulations with background magnetic fields parallel to the temperature gradient and $\beta_{e0}=120$. $L_{\mathrm{T}0} / \rho_{e0}$ is varied between 44 and 150 in order to increase $\beta_{e0} \rho_{e0} / L_{\mathrm{T}0}$ past unity without changing $\beta_{e0}$. The heat flux is normalised to the free-streaming value. A black line is drawn at $\langle \delta B^2 / B_0^2 \rangle = 1$ to indicate when the perturbed magnetic energy exceeds that of the background field. }
\label{fig:2dLT_time_evol}
\end{figure}

\begin{figure}
    \centering
    \includegraphics[width=0.49\linewidth]{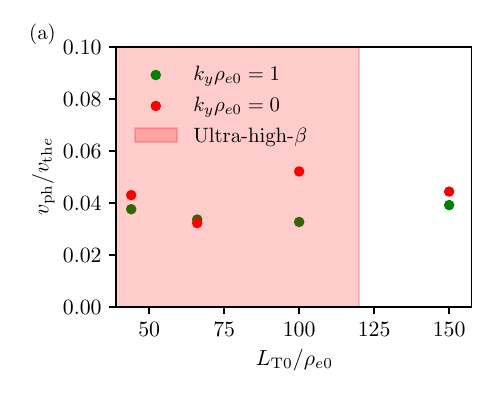}
    \includegraphics[width=0.49\linewidth]{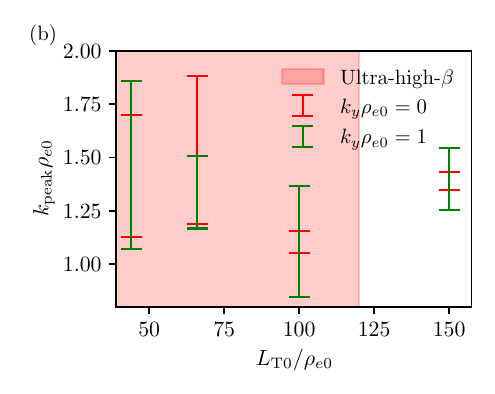}
    \caption{Values of a) $v_\mathrm{ph} / v_{\mathrm{th}e}$ and b) the peak wavenumber for simulations with $\beta_{e0}=120$ and varying $L_{\mathrm{T}0} / \rho_{e0}$ for both parallel ($k_y \rho_{e0} = 0$) and oblique ($k_y \rho_{e0} = 1$) modes. The peak wavenumber was found for each time step in the last $350 \Omega_{e0}^{-1}$ of each simulation and the mean of these values is plotted. Error bars show the standard deviation of this value.}
    \label{fig:2dLT_fourier}
\end{figure}

To investigate whether the saturation state of the WHFI depends solely on $\beta_{e0}$ in the ultra-high-$\beta$ regime, we ran simulations with fixed $\beta_{e0}=120$ and varied $L_{\mathrm{T}0} / \rho_{e0}$. Since a diffusive heat-flux component would depend on the temperature-gradient length-scale, this provides a test of the extent to which  diffusive heat transport contributes to the total heat flux.

The temporal evolution of $\langle \delta B^2 /B_0^2 \rangle$ and $\langle q_{e\parallel} / q_{\mathrm{fs}} \rangle$ is plotted in figure \ref{fig:2dLT_time_evol}. This figure shows that, for $\beta_{e0} \rho_{e0} / L_{\mathrm{T}0} \gtrsim 1$, the behaviour of each simulation is largely independent of $L_{\mathrm{T}0} / \rho_{e0}$. All simulations reach saturation at similar times with similar values of $\langle \delta B^2 /B_0^2 \rangle$ and $\langle q_{e\parallel} / q_{\mathrm{fs}} \rangle$. The only simulation that differs slightly has $L_{\mathrm{T}0} / \rho_{e0} = 150$, and hence is not in the ultra-high-$\beta$ regime, so is not expected to exhibit the same behaviour. Although $q_{e\parallel }/q_{\mathrm{fs}}$ is independent of $L_{\mathrm{T}}/{\rho }_{e0}$ in both ultra-high and moderately-high-$\beta$ regimes, there is no \textit{a priori} reason to expect it to saturate at the same value in both. Therefore, the slightly lower saturated heat flux in the $L_{\mathrm{T}0} / \rho_{e0} = 150$ simulation is not entirely unexpected. The approximate independence of $\langle \delta B^2 /B_0^2 \rangle$ and $\langle q_{e\parallel} / q_{\mathrm{fs}} \rangle$ from the temperature-gradient length scale is indicative of a change in scaling from the moderately-high-$\beta$ regime, where the saturation value of $\langle \delta B^2 / B_0^2 \rangle$ depends on $L_{\mathrm{T}0} / \rho_{e0}$. Thus, these results suggest that diffusive contributions to heat transport are minimal. We note that the initial peak in the heat flux at $t \Omega_{e0} < 100$ occurs at different times for the different simulations, because changing the simulation box width changes the time required for the initial heat flux to be established.

We also find that $L_{\mathrm{T}0} / \rho_{e0}$ does not significantly affect the dispersion relation or peak wavenumber of the whistler waves. This is shown in figure \ref{fig:2dLT_fourier}, where $v_\mathrm{ph} / v_{\mathrm{th}e}$ and the peak wavenumber are both approximately independent of $L_{\mathrm{T}0} / \rho_{e0}$. This again agrees with the expectations of \S \ref{sec:large_whistler_advective} and indicates that the phase velocity depends primarily on $\beta_{e0}$. 

\subsubsection{Cross-field transport} \label{sec:angled_results}

\begin{figure}
    \centering
    \includegraphics[width=\linewidth]{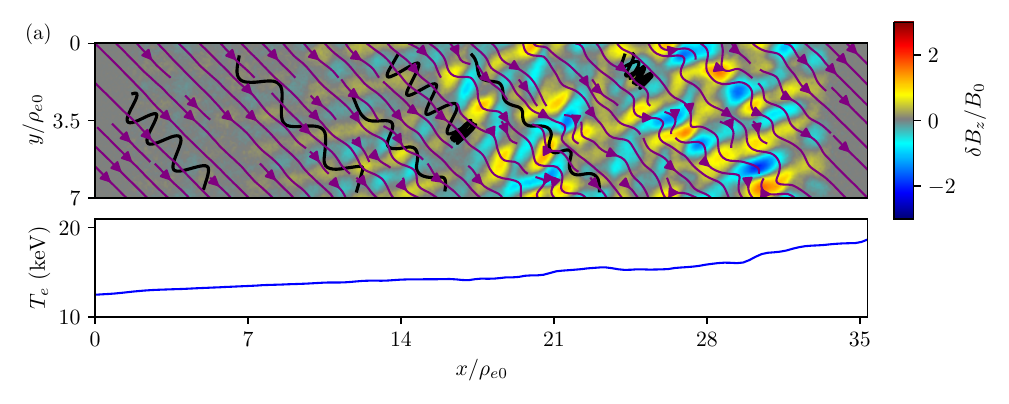}
    \includegraphics[width=\linewidth]{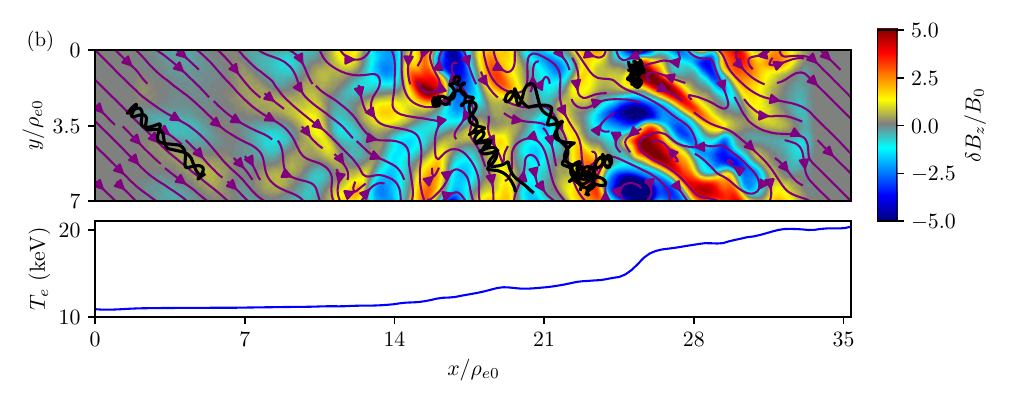}
    \caption{Magnetic field and $T_e$ at a) $t\Omega_{e0} = 54$, during exponential wave growth, and b) $t\Omega_{e0} = 570$, during nonlinear growth. Images are taken from a 2D3V simulation with an oblique background magnetic field and $\beta_{e0} = 400$. Purple lines show the magnetic field lines in the plane of view, while the colour plot shows the out-of-plane component. Black lines show test-electron trajectories sampled around the times of the field snapshots.
}
    \label{fig:angled_waves}
\end{figure}

\begin{figure}
    \centering
    \includegraphics[width=\linewidth]{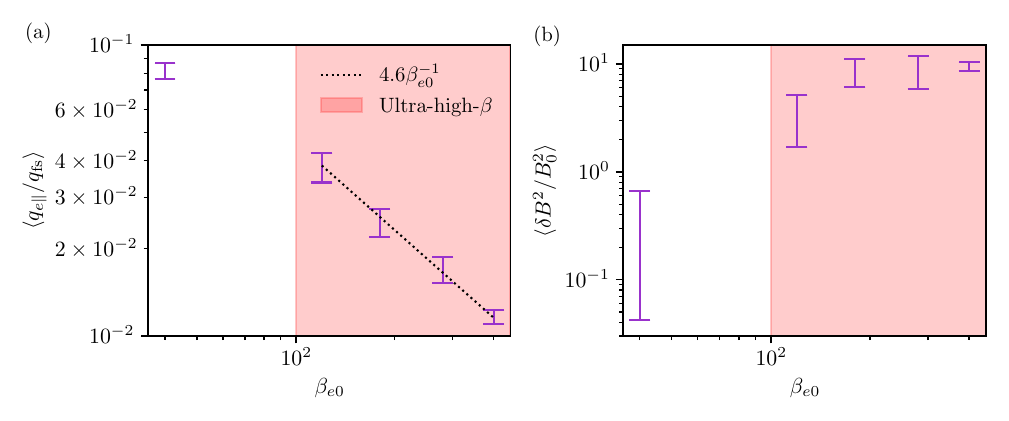}
    \caption{Scalings of (a) the heat flux and (b) the perturbed magnetic energy with $\beta_{e0}$ in 2D3V simulations with $L_{\mathrm{T}0} / \rho_{e0}=100$ and angled magnetic fields, where the values on the $y$ axes have been spatially averaged over the small-$L_\mathrm{T}$ region and temporally averaged over the saturated interval. In (a), a linear best-fit of the form $\langle q_{e\parallel}/ q_{\mathrm{fs}} \rangle = A\beta_{e0}^{-1}$, where $A \approx 4.6$ is a fitting parameter, is shown by the black dotted line in the ultra-high-$\beta$ region.}
    \label{fig:ang_scaling}
\end{figure}

If the diffusive heat-flux component is indeed small compared to the advective component, perpendicular transport should remain strongly suppressed. To investigate whether the heat flux remains anisotropic for large-amplitude whistler waves, 2D3V simulations were initialised with conditions similar to those in \S \ref{sec:2d_beta_results} but with the magnetic field inclined at $45^{\circ}$ to the temperature gradient as described in table \ref{tab:simulations}. This produces a diamagnetic heat flux through the Righi-Leduc effect and introduces a component of the temperature gradient perpendicular to the magnetic field, as in \citet{choudhury_modeling_2025}. To keep $L_{\mathrm{T}0} / \rho_{e0} = 100$, the thermal bath boundary conditions were kept at the same temperatures and the dimensions of the box were changed to $L_x = 35 \rho_{e0}$ and $L_y = 7.1 \rho_{e0}$. Example magnetic fields during exponential whistler growth and nonlinear whistler growth are shown in figure \ref{fig:angled_waves} for $\beta_{e0} = 400$, showing how the whistler waves propagate along the magnetic field lines, at a $45^{\circ}$ angle to the temperature gradient. The figure shows locally closed or strongly distorted field-line structures and magnetic mirrors, which effectively trap and reflect electrons. The electron trajectories remain closely tied to the  magnetic-field lines, suggesting that perpendicular diffusion is weak.

$\langle \delta B^2/B_0^2 \rangle$ and $\langle q_{e\parallel}/q_{\mathrm{fs}} \rangle$ scale with $\beta_{e0}$ similarly to the case with magnetic fields applied parallel to the temperature gradient, as shown in figure \ref{fig:ang_scaling}. The beginning of the plateau in $\langle \delta B^2/B_0^2 \rangle$ seen in figure \ref{fig:2d_scaling} is found to continue, suggesting that the instability saturates with $\delta B \sim B_0$, and the best fit to the heat-flux scaling of $q_{e\parallel}/q_{\mathrm{fs}} \sim 4.6\beta_{e0}^{-1}$ is very similar to that seen in figure \ref{fig:2d_scaling}. The similarities show that the scaling relationships observed in figure \ref{fig:2d_scaling} and predicted in \S \ref{sec:large_whistler_advective} continue far into the ultra-high-$\beta$ regime. In an advective heat-flux model, the parallel heat-flux and magnetic-energy scalings should not depend strongly on whether the background magnetic field is parallel or oblique to the temperature gradient. The agreement between the two sets of simulations therefore supports this model.

\begin{figure}
    \centering
    \includegraphics[width=\linewidth]{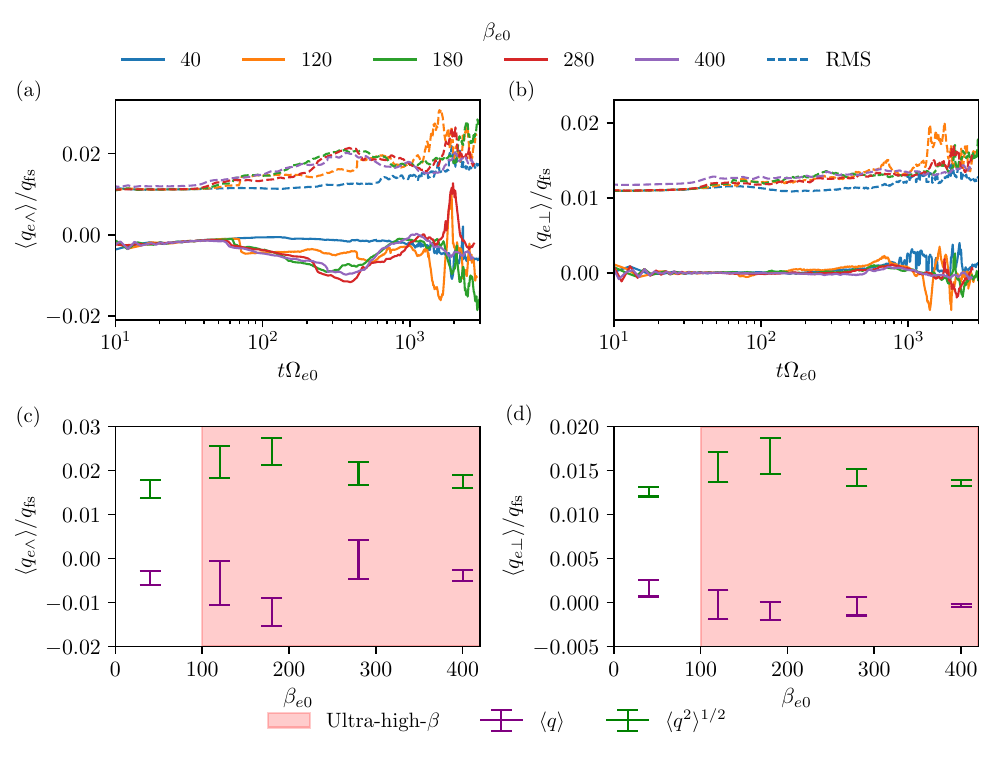}
    \caption{Temporal evolution of the spatially averaged (solid lines) and RMS (dashed lines) (a) $q_{e\wedge} $ and (b) $q_{e\perp}$, normalised to $q_\mathrm{fs}$. Averages are taken over the small-$L_\mathrm{T}$ region. Panels (c) and (d) show how the cross-field transport, temporally averaged over the saturated interval, varies with $\beta_{e0}$, where $1\sigma$ (standard deviation of the spatially averaged value over time steps in the saturated interval) error bars are plotted.}
    \label{fig:angled_trends}
\end{figure}

The temporal evolution of $q_{e\wedge} / q_\mathrm{fs}$ and $q_{e\perp} / q_\mathrm{fs}$, as well as their RMS values, are plotted in figure \ref{fig:angled_trends}. Saturation is reached after around $1000 \Omega_{e0}^{-1}$ for each of the simulations. The average and RMS cross-field transport values during the saturated interval are calculated and also plotted in figure \ref{fig:angled_trends}. As the WHFI grows, the mean and RMS diamagnetic heat flux increase in the ultra-high-$\beta$ simulations. However, as the instability approaches saturation, the magnitude of the mean diamagnetic heat flux decreases again and begins to fluctuate strongly. The magnitude of these fluctuations decreases with increasing $\beta_{e0}$ for $\beta_{e0}>180$, such that the highest-$\beta_{e0}$ simulations had relatively small mean and RMS diamagnetic fluxes. The mean perpendicular heat flux remains very small throughout the simulation. This suggests that the initial increase in the mean diamagnetic heat flux before saturation is due to the Righi-Leduc effect, which is then suppressed once the whistler waves become sufficiently large. As there remains a non-zero diamagnetic heat flux due to the Righi-Leduc effect, it is possible that at sufficiently high $\beta_{e0}$ the diamagnetic heat flux could become larger than the parallel heat flux.

The reason the RMS heat flux is consistently so much higher than the mean cross-field heat flux can be seen clearly in figure \ref{fig:angled_waves}. The electrons remain largely confined to moving along the magnetic-field lines. However these field lines undergo large deviations from their initial direction, causing the electrons to follow the crests and troughs of the whistler perturbations. This means that at any given time, an electron is likely to be travelling at an oblique angle to the initial external field, but all of these fluctuations cancel out, leading to very small mean perpendicular transport. The fact that the mean perpendicular transport is very small regardless of $\beta_{e0}$ rules out a diffusive, isotropic heat-flux model, especially given the perpendicular direction has a component down the temperature gradient. This behaviour is consistent with the theory outlined in \S \ref{sec:large_whistler_advective}: the WHFI strongly affects parallel heat transport while leaving perpendicular transport suppressed.

\subsection{1D3V results}\label{sec:1d_results}

Since the whistler waves in the ultra-high-$\beta$ regime are large-amplitude fluctuations, scattering of electrons is no longer expected to depend solely on resonant processes. Therefore, the requirement that oblique waves be present to resonantly scatter heat-flux-carrying electrons no longer applies, and it is plausible that heat flux suppression in 1D3V simulations may be observed. We have therefore run 1D3V simulations similar to those described in \S \ref{sec:2d_results} to compare the results with the 2D3V case and test whether they also follow the advective heat-flux model from \S \ref{sec:large_whistler_advective}.

\subsubsection{Varying plasma beta} \label{sec:1d_beta_results}
\begin{figure}
    \centering
    \includegraphics[width=\linewidth]{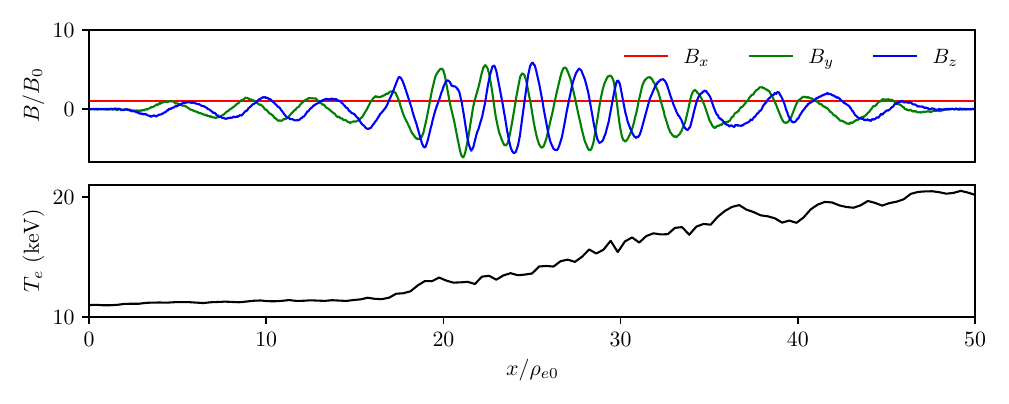}
    \caption{The magnetic fields and $T_e$ for a 1D3V simulation with $\beta_{e0} = 280$ and $L_{\mathrm{T}0} / \rho_{e0} = 100$ when the WHFI is undergoing nonlinear growth at $t\Omega_{e0} = 1617$. The temperature increases along the $x$ direction, and the background magnetic field is parallel to the $x$ axis.
}
    \label{fig:1d_waves}
\end{figure}

\begin{figure}
    \centering
    \includegraphics[width=0.7\linewidth]{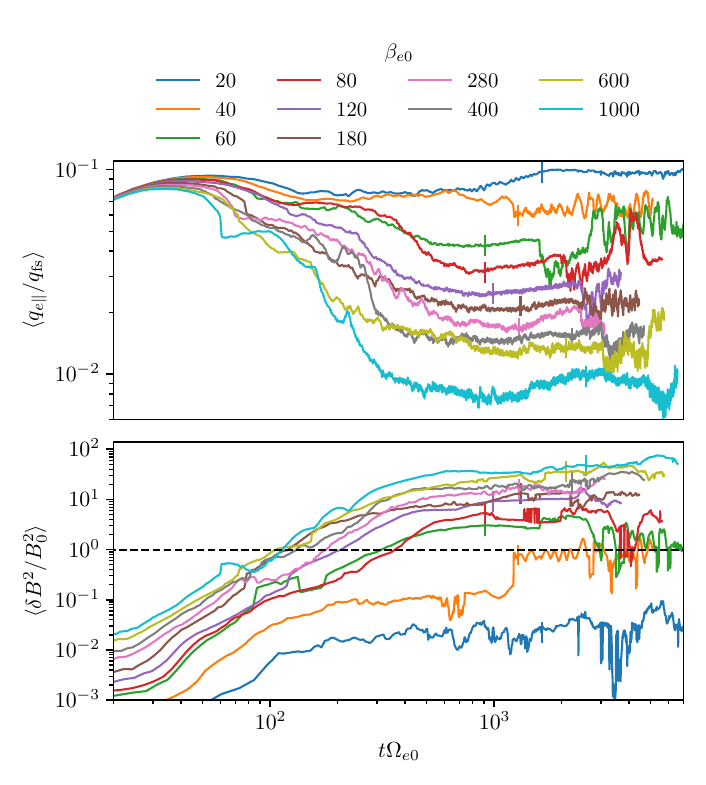}
    \caption{Temporal evolution of the spatially averaged parallel heat flux (upper) and perturbed magnetic energy (lower) for 1D3V simulations with background magnetic fields parallel to the temperature gradient and at a range of $\beta_{e0}$ values. The heat flux is normalised to the free-streaming value. Both quantities are averaged across the small-$L_\mathrm{T}$ region. A black line is drawn at $\langle \delta B^2 / B_0^2 \rangle= 1$ to show when the perturbed magnetic energy exceeds that of the background field and markers are drawn on each of the lines to show where saturation was determined to have been reached (the start of the saturated interval). Particularly sharp gradients are typically caused by rapid changes in the width of the small-$L_\mathrm{T}$ region.}
    \label{fig:1d_time_evol}
\end{figure}

1D3V simulations were run with varying plasma beta, similar to those in \S \ref{sec:2d_beta_results}. Because 1D3V simulations are computationally simpler, we were able to investigate higher values of $\beta_{e0}$, namely $\beta_{e0} \in \{40, 60, 80, 120, 180, 280, 400, 600, 1000\}$ (as described in table \ref{tab:simulations}). An example of the magnetic field and electron temperature data is shown in figure \ref{fig:1d_waves}, showing the formation of whistler waves and a steep temperature gradient in the same spatial region, similar to that observed in the 2D3V simulations.

\begin{figure}
    \centering
    \includegraphics[width=0.7\linewidth]{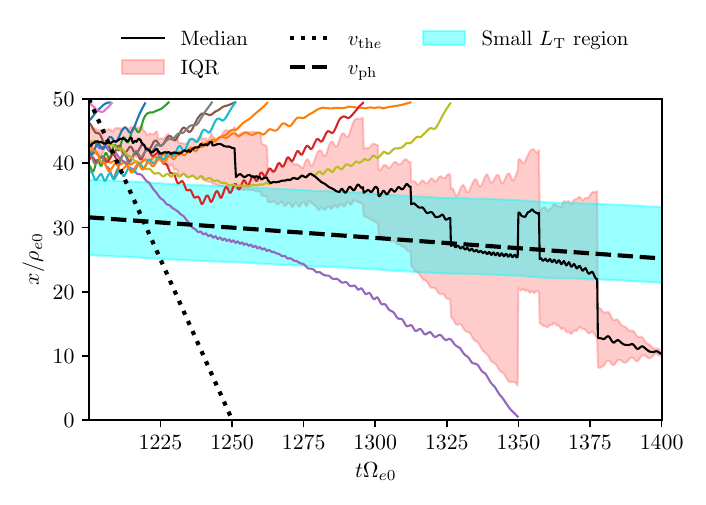}
    \caption{Electron trajectories for particles that were on the hot side of the small-$L_\mathrm{T}$ region at $t\Omega_{e0}=1200$ in the 1D3V simulation with $\beta_{e0}=120$ and $L_{\mathrm{T}0} / \rho_{e0} = 100$. Only a subset of electrons satisfying these criteria are plotted to avoid overcrowding the figure (coloured lines); however, the plotted median and IQR of electron positions are computed using all electrons that met the criteria. The median and IQR are highly variable due to the smaller number of tracked particles in 1D3V compared to 2D3V. The position of the small-$L_\mathrm{T}$ region at each time is shown in blue. The thermal velocity and whistler phase velocity $v_\mathrm{ph}$ are shown for reference. }
    \label{fig:1d_trajs}
\end{figure}

Figure \ref{fig:1d_time_evol} shows that very strong amplification of the perturbed magnetic fields ($\delta B ^2 / B_0^2 \gg 1$) is seen in 1D3V, leading to significant suppression of parallel heat flux despite the absence of oblique whistler modes. The ability of parallel modes to affect both co- and counter-propagating electrons is confirmed by the trajectories of individual PIC macroparticles shown in figure \ref{fig:1d_trajs}. Electrons travelling both towards the hot wall and towards the cold wall reverse direction on timescales of tens of $\Omega_{e0}^{-1}$, which is impossible in the small-amplitude-whistler regime where only particles travelling towards the hot wall can be resonantly scattered. Therefore, there must be other mechanisms that act to alter electron trajectories in this large-amplitude-whistler regime, suggesting the theory outlined in \S \ref{sec:large_whistler_advective} may apply to 1D3V simulations as well. 

Figure \ref{fig:1d_trajs} confirms that the small-$L_\mathrm{T}$ region acts as a transport barrier in 1D3V, as in 2D3V. Very few electrons pass through the small-$L_\mathrm{T}$ region at all, with the majority being reflected back towards the hot wall. The electrons that make it through the transport barrier typically interact strongly with it, travelling much slower than the thermal velocity when inside the barrier. The fact that some electrons are able to pass through the barrier suggests that some diffusion occurs; however, the fact that this occurs for only a small fraction of the electrons is consistent with the assumption made in \S \ref{sec:large_whistler_advective} that diffusion should be negligible.

\begin{figure}
    \centering
    \includegraphics[width=0.49\linewidth]{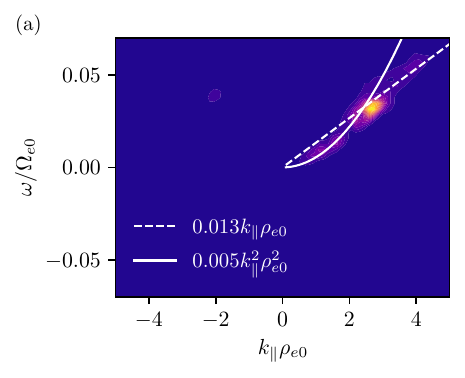}
    \includegraphics[width=0.49\linewidth]{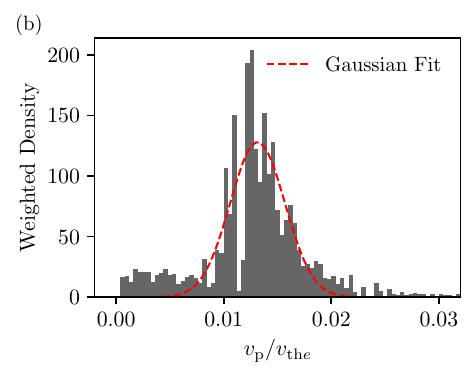}
    \caption{(a) Dispersion relation for a 1D3V simulation with $\beta_{e0}=280$ and $L_{\mathrm{T}0} / \rho_{e0}=100$, computed using waves during the last $2000 \Omega_{e0}^{-1}$ of simulation time during the saturated interval and excluding data in the masking region. Two best-fitting curves have been plotted: a linear dashed line and a quadratic solid line. b) a histogram showing the probability density of different phase velocities for the same simulation, weighted by the power at each phase velocity (i.e. the brightness in a))
    }
    \label{fig:1dbeta280_dispersion}
\end{figure}

To determine whether the heat flux in 1D3V simulations is advective, we calculate the whistler phase velocity. The dispersion relation of whistler waves in the 1D3V simulations differs from the 2D3V simulations, leading to potential differences in heat flux between the two cases. This can be seen in figure \ref{fig:1dbeta280_dispersion}, where the dispersion relation for the simulation with $\beta_{e0}=280$ has been plotted. The dispersion of the waves has a quadratic dependence on $k_{\parallel} \rho_{e0}$ when $k_\parallel \rho_{e0} < k_{\parallel\mathrm{peak}} \rho_{e0}$, and a linear dependence when $k_\parallel \rho_{e0} > k_{\parallel\mathrm{peak}} \rho_{e0}$, where the peak wavenumber $k_{\parallel\mathrm{peak}} \rho_{e0}$ is easily identifiable by the region of maximum power. Thus the dispersion relation is similar to the small-amplitude regime at small wavenumbers, and to the 2D3V large-amplitude regime at large wavenumbers.

The impact of the two different dispersion regimes on the phase velocity is shown in figure \ref{fig:1dbeta280_dispersion}b, where the power-weighted probability density of different phase velocities is plotted. It can be seen that the phase velocities are sharply peaked around one value, and that the presence of two different dispersion relations does not result in two different phase velocities being present. This is confirmed by noting that the region of highest power in figure \ref{fig:1dbeta280_dispersion}a is near the intersection of the linear and quadratic fits. A Gaussian was fitted to the probability-density distribution of phase velocities, the mean of which is taken to be the phase velocity for whistler waves in this simulation. The same method was used for all 1D3V simulations to find their phase velocities. 

\begin{figure}
    \centering
    \includegraphics[width=0.49\linewidth]{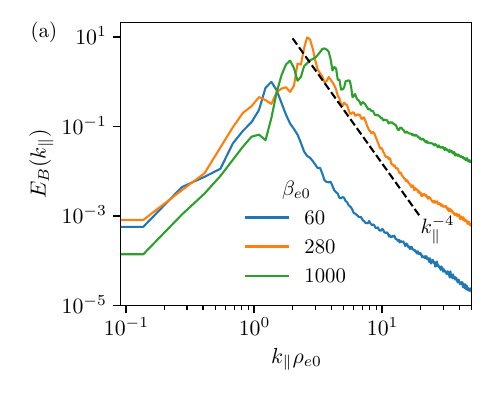}
    \includegraphics[width=0.49\linewidth]{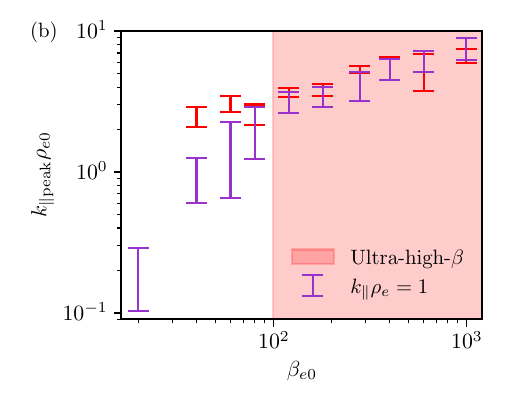}
    \caption{(a) Fourier spectra of $B_y + \mathrm{i}B_z$ for 1D3V simulations with $\beta_{e0} \in \{60,280, 1000\}$ and $L_{\mathrm{T}0} / \rho_{e0}=100$, averaged over the last $2000 \Omega_{e0}^{-1}$ of simulation time during the saturated interval and excluding data in the masking region. The spectra have been normalised such that $\int \mathrm{d}k_\parallel \ E_B (k_\parallel) = \langle \delta B_y^2 +\delta B_z^2 \rangle / B_0^2$. The $k_\parallel^{-4}$ scaling typical of whistler waves is shown over its relevant region. (b) Plots of how the peak wavenumber (red) scales with $\beta_{e0}$, and how this compares with its expected values if the assumption $k_\parallel \rho_e \sim 1$ is valid. Data are taken for 1D3V simulations with $L_{\mathrm{T}0}/ \rho_{e0}=100$. If $k_\parallel \rho_e \sim 1$, then $k_\parallel \rho_{e0} \sim \delta B / B_0$, hence the actual quantity plotted with the purple error bars is $\sqrt{\langle \delta B^2 / B_0^2\rangle}$ as an estimate of this. All values have been averaged over the saturated interval and $\sqrt{\langle \delta B^2 / B_0^2\rangle}$ was spatially averaged over the small-$L_\mathrm{T}$ region. The error bars come from the standard deviation of these quantities during the saturated interval (only the last $2000 \Omega_{e0}^{-1}$ of simulation time for the peak wavenumber).
    }
    \label{fig:1d_peak_k_scaling}
\end{figure}

Because in 1D the perturbed magnetic field continues to grow beyond order-unity amplitudes, it is also possible to test the assumption that the peak wavenumber follows $k \rho_{e} \sim k \rho_{e0} B_0 / \delta B \sim 1$. The Fourier spectra of whistler waves in 1D3V simulations, averaged over the final $2000 \Omega_{e0}^{-1}$ of the simulation, are plotted in figure \ref{fig:1d_peak_k_scaling}a. The spectra remain similar to those in both 2D3V and small-amplitude simulations, following $k_\parallel^{-4}$ for $k_\parallel \rho_{e0} \lesssim 10$, but the peak wavenumber increases with $\beta_{e0}$. To estimate $\delta B / B_0$,  we plot $\sqrt{\langle \delta B^2 / B_0^2\rangle}$ alongside the peak $k \rho_{e0}$ in figure \ref{fig:1d_peak_k_scaling}b. The peak wavenumber was found using the same method as in \S \ref{sec:2d_beta_results}. It can be seen that once $\sqrt{\langle \delta B^2 / B_0^2\rangle} \gtrsim 2$, the assumption that $k \rho_{e} \sim k \rho_{e0} B_0 / \delta B \sim 1$ applies well, again validating the proposed theory.

\begin{figure}
    \centering
    \includegraphics[width=\linewidth]{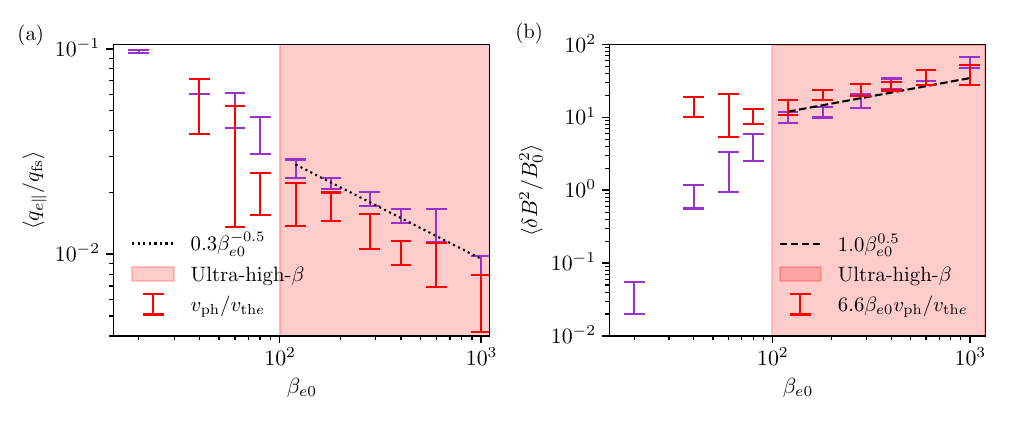}
    \caption{Scalings of (a) the heat flux and (b) the perturbed magnetic energy with $\beta_{e0}$ in 1D3V simulations for $L_{\mathrm{T}0} / \rho_{e0}=100$, where the values on the $y$ axes have been spatially averaged over the small-$L_\mathrm{T}$ region and temporally averaged over the saturated interval. A black dashed best-fitting line to the heat flux values is plotted. The expected values for both the perturbed magnetic energy and heat flux from the measured phase velocity are plotted, (multiplied by an order-unity constant for the perturbed magnetic energy). A best-fitting scaling relation for the perturbed magnetic energy has been plotted, based on the best-fitting scaling relation plotted in (a) and (\ref{eq:B_predict_alpha}). The phase velocities and $1\sigma$ error bars are calculated from the Gaussian fit to the phase velocity probability density function, as shown in figure \ref{fig:1dbeta280_dispersion}(b).
    }
    \label{fig:1d_scaling}
\end{figure}

Figure \ref{fig:1d_scaling} shows the scalings for $\langle \delta B^2 / B_0^2 \rangle$ and $\langle q_{e\parallel} / q_{\mathrm{fs}} \rangle$ with $\beta_{e0}$. $\langle \delta B^2 / B_0^2 \rangle$ and $\langle q_{e\parallel} / q_{\mathrm{fs}} \rangle$ behave differently from their 2D3V counterparts, with $\langle \delta B^2 / B_0^2 \rangle$ continuing to increase past $\delta B \sim B_0$. Furthermore, $\langle q_{e\parallel} / q_{\mathrm{fs}} \rangle \sim \beta_{e0}^{-0.5}$, indicating weaker suppression than in 2D. Although the scaling is different from the 2D3V simulations, the results follow the proposed advective theory well. The heat flux and $v_\mathrm{ph} / v_{\mathrm{th}e}$ agree reasonably well; their scalings with $\beta_{e0}$ are very similar, and they quantitatively differ by only ${\sim}$30\%. The scaling $\delta B^2 / B_0^2 \sim \beta_{e0} v_\mathrm{ph} / v_{\mathrm{th}e}$ appears to hold within an order-unity factor as well. This supports the notion that the advective model does indeed apply in 1D3V simulations. A further aspect of figure \ref{fig:1d_scaling} to note is the very steep drop-off in $\delta B^2 / B_0^2$ and enhancement of the heat flux for $\beta_{e0}<100$, corresponding to simulations that are not in the ultra-high-$\beta$ regime. This helps explain why significant heat-flux suppression in 1D3V simulations has not been observed in prior studies, as it only operates in the ultra-high-$\beta$ regime.

\subsubsection{Varying temperature gradient} \label{sec:1dLTresults}

\begin{figure}
    \centering
    \includegraphics[width=0.7\linewidth]{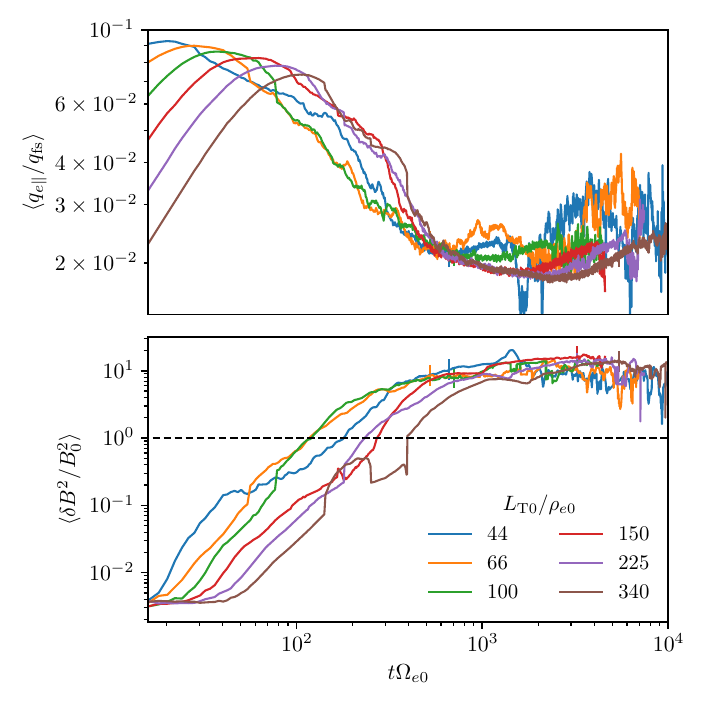}
    \caption{Temporal evolution of the spatially averaged parallel heat flux (upper) and perturbed magnetic energy (lower) for 1D3V simulations with background magnetic fields parallel to the temperature gradient and $\beta_{e0}=180$. $L_{\mathrm{T}0} / \rho_{e0}$ is varied between 44 and 340 in order to vary $\beta_{e0} \rho_{e0} / L_{\mathrm{T}0}$ without changing $\beta_{e0}$. The heat flux is normalised to the free-streaming value. A black line is drawn at $\langle \delta B^2 / B_0^2 \rangle= 1$ to show when the perturbed magnetic energy exceeds that of the background field.}
    \label{fig:1d_LT_time_evol}
\end{figure}

\begin{figure}
    \centering
    \includegraphics[width=0.49\linewidth]{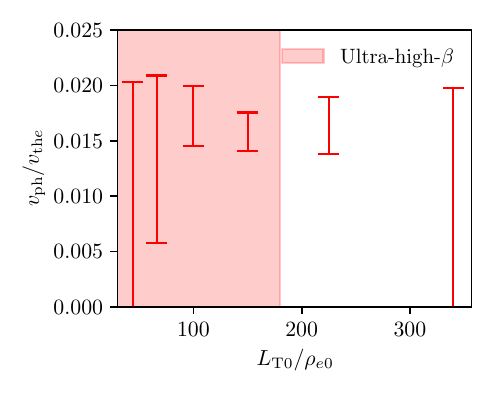}
    \caption{Values of $v_\mathrm{ph} / v_{\mathrm{th}e}$ for simulations with $\beta_{e0}=180$ and varying $L_{\mathrm{T}0} / \rho_{e0}$. The mean phase velocity and $1\sigma$ error bars were calculated in the same way as in figure \ref{fig:1d_scaling}.}
    \label{fig:1dLT_disp}
\end{figure}

As in \S \ref{sec:2dLTresults}, the transition into the ultra-high-$\beta$ regime was also studied in 1D3V by keeping $\beta_{e0}$ constant and varying $L_{\mathrm{T}0} / \rho_{e0}$. Because 1D3V simulations are computationally cheaper, $\beta_{e0}=180$ was used rather than $\beta_{e0}=120$. The temporal evolution of both the heat flux and perturbed magnetic energy is plotted in figure \ref{fig:1d_LT_time_evol}. As in the 2D3V case, all simulations behave very similarly, suggesting that the saturated WHFI becomes independent of $L_{\mathrm{T}0}$ in this regime.

Likewise, the dispersion relation appears to be independent of $L_{\mathrm{T}0} / \rho_{e0}$ in both the linear and quadratic regions, as shown in figure \ref{fig:1dLT_disp}. The phase velocity remains constant within the uncertainty, regardless of $L_{\mathrm{T}0} / \rho_{e0}$, as expected for a quantity that depends only on $\beta_{e0}$.

\section{Discussion and conclusions}\label{sec:discussion}

\subsection{Key findings} \label{sec:conclusion}

In this paper, we have developed and tested a theory for the WHFI in ultra-high-$\beta$ plasmas, defined by $\beta_{e0}\rho_{e0}/L_{\mathrm{T}0}\gtrsim1$. In this regime, extrapolation of the moderately-high-$\beta$ theory predicts magnetic fluctuations with $\delta B/B_0\gtrsim1$, so the usual quasilinear, small-amplitude description is no longer self-consistent. We therefore proposed a large-amplitude saturation picture and heuristically derived two possible heat-flux closures: a diffusive model (\S \ref{sec:large_whistler_diffusive}) and an advective model (\S \ref{sec:large_whistler_advective}). Comparison with 2D3V and 1D3V PIC simulations shows that the advective model describes the simulations, whereas the diffusive model does not. Our main results are as follows:
\begin{enumerate}
    \item In 2D3V simulations, the magnetic energy in the whistler fluctuations can become larger than that in the background field but only by an order-unity factor, such that $\delta B \sim B_0$ is a reasonable approximation of the saturated state. Conversely, in 1D3V the amplification continues to increase over the range of $\beta_{e0}$ investigated. 

    \item The parallel heat flux is strongly suppressed relative to the free-streaming value. In 2D3V, we find

    \begin{equation}
        q_{e\parallel}  \approx 4.7 \beta_{e0}^{-1} q_{\mathrm{fs}}  \approx 2.65 \beta_{e0}^{-1} n_{e0} v_{\mathrm{th}e0}T_{e0},
    \end{equation}
    while in 1D3V the suppression is weaker, approximately following $q_{e\parallel}/q_{\mathrm{fs}}\approx0.3 \beta_{e0}^{-1/2}$.

    \item The heat flux is controlled by the whistler phase velocity, with $q_{e\parallel}/q_{\mathrm{fs}}\sim v_{\mathrm{ph}}/v_{\mathrm{th}e}$, as predicted by the advective model.

    \item The large-amplitude whistler fluctuations collectively form an effective transport barrier that traps or reflects most heat-carrying electrons. Thermal energy is therefore transported down the temperature gradient at approximately the whistler phase velocity rather than by spatial diffusion.

    \item Perpendicular transport remains strongly suppressed, even when the imposed temperature gradient has a component perpendicular to the background magnetic field. This is inconsistent with an isotropic diffusive heat-flux model.
    \item In the ultra-high-$\beta$ regime, the saturated heat flux and whistler phase velocity are approximately independent of $L_{\mathrm{T}0}/\rho_{e0}$ at fixed $\beta_{e0}$ within our numerical uncertainties, which is again consistent with the advective interpretation. 
\end{enumerate}

The finding that the WHFI-regulated heat flux is primarily advective in ultra-high-$\beta$ plasmas, rather than diffusive, merits further discussion. The diffusive heat-flux model assumes that the electrons carrying most of the heat flux could undergo a random walk through the large-amplitude whistler fluctuations. Our simulations show otherwise; instead, their trajectories are such that the majority do not cross the transport barrier. Some electrons are deflected as they stream into the region of strongest fields, and subsequently stream away from the barrier. Others become trapped in local magnetic mirrors or locally closed-field line configurations. 
This implies that once the fluctuations become order-unity in amplitude, the transport mechanism changes: heat-flux suppression is no longer controlled primarily by quasilinear pitch-angle scattering, but by the motion of large-amplitude whistler structures that act as barriers to electron transport. Although an advective picture of WHFI-moderated heat transport has also been proposed for plasmas in the moderately-high-$\beta$ regime by \citet{roberg-clark2018} and \citet{drake_whistler-regulated_2021}, that picture differs substantively from the advective transport here. In those models, heat transport relies on cyclotron-resonant scattering of electrons by mobile scattering centres propagating at the whistler phase speed, a distinctive mechanism from the motion of large-amplitude structures that arise in the ultra-high-$\beta$ regime considered here. In short, our results show that the transition to the ultra-high-$\beta$ regime is not merely a quantitative extension of the moderately-high-$\beta$ WHFI, but instead involves a qualitative change. 

\subsection{Differences between 1D3V and 2D3V} \label{sec:disc:diffs}

\begin{figure}
    \centering
    \includegraphics[width=\linewidth]{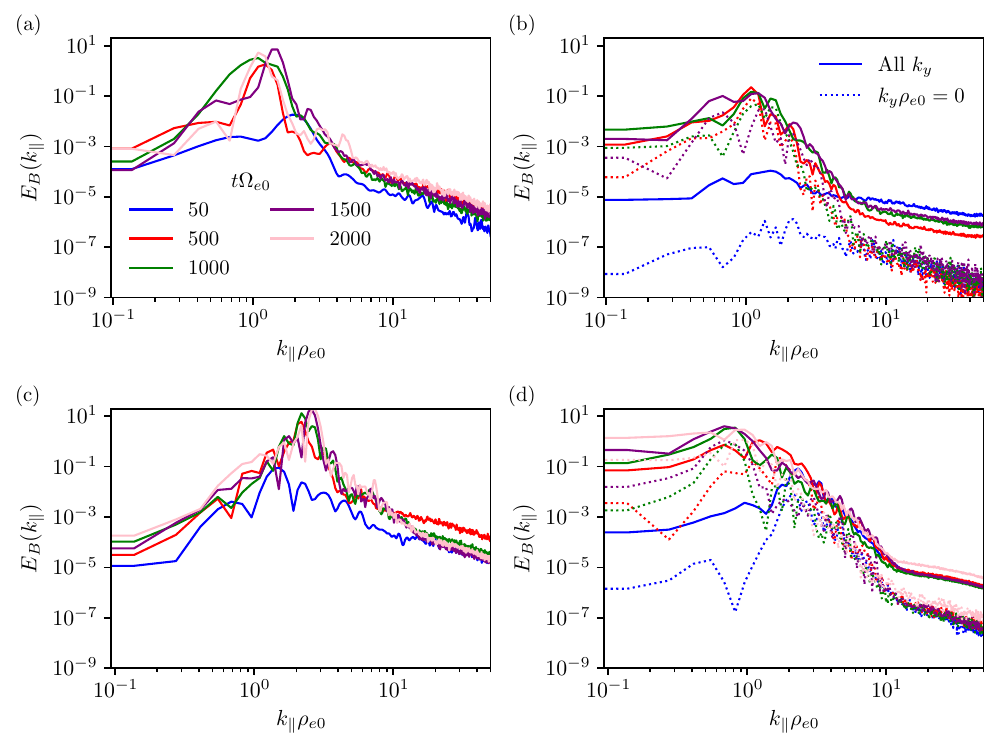}
    \caption{Fourier spectra of $B_y + \mathrm{i}B_z$ at different times for (a) 1D3V $\beta_{e0}=60$, (b) 2D3V $\beta_{e0}=60$, (c) 1D3V $\beta_{e0}=280$, and (d) 2D3V $\beta_{e0}=280$ simulations. For the 1D3V plots, the Fourier spectra have been averaged over $100 \Omega_{e0}^{-1}$ around the time shown. For the 2D3V plots, the average was taken over $40 \Omega_{e0}^{-1}$ and the Fourier spectra are for both the parallel modes and for the spectra integrated over all $k_y$. No Fourier spectrum is shown in panel (b) for $t \Omega_{e0} = 2000$, since the simulation did not run for that long. Fourier spectra have been normalised such that the integral of each spectrum equals $\langle \delta B_y^2 + \delta B_z^2 \rangle / B_0 ^2$, the perturbed magnetic energy perpendicular to the background magnetic field in the relevant modes at that time. All plots have the same axes limits for ease of comparison.}
    \label{fig:timespectra}
\end{figure}

\begin{figure}
    \centering
    \includegraphics[width=0.49\linewidth]{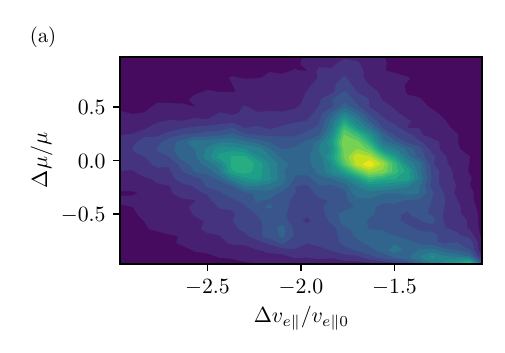}
    \includegraphics[width=0.49\linewidth]{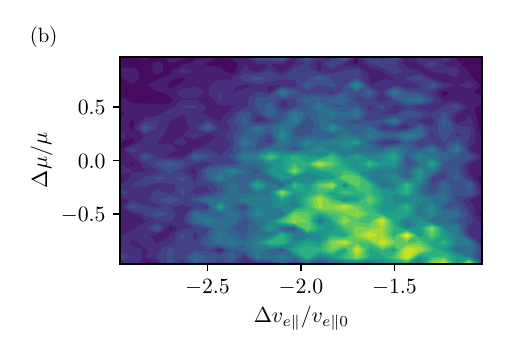}
    \caption{Probability-density maps for the relative changes in magnetic moment and velocity for particles that undergo reflection events, as defined in the text, in both (a) 1D3V and (b) 2D3V simulations. Both simulations have $\beta_{e0}=120$. In 1D3V, the magnetic moment is approximately conserved during these reflection events, whereas in 2D3V it is not.}
    \label{fig:mag_moment}
\end{figure}

Although the advective model captures the heat flux in both 1D3V and 2D3V simulations when the measured whistler phase velocity is used, the two geometries exhibit different dispersion relations, peak wavenumbers, and scalings with $\beta_{e0}$. This indicates that the dimensionality affects the nonlinear evolution of the whistler spectrum, even though the behaviour of the heat flux remains consistent with the advective model. Because the whistler waves are large-amplitude in the ultra-high-$\beta$ regime, these differences are likely associated with nonlinear interactions among whistler modes, which need not proceed in the same way in 1D3V and 2D3V.

To interpret the difference in behaviour between the two types of simulations, Fourier spectra at different times during the simulation for different values of $\beta_{e0}$ in both 1D3V and 2D3V are shown in figure \ref{fig:timespectra}. Several key differences are apparent. First, figure \ref{fig:timespectra} shows that in 1D, the peak wavenumber becomes larger for higher $\beta_{e0}$, which is not observed in 2D3V. This is related to the fact that the perturbed magnetic energy continues to increase with $\beta_{e0}$ past $\beta_{e0} \rho_{e0} / L_{\mathrm{T}0} \sim 1$ in 1D3V, decreasing the local electron Larmor radius $\rho_e$ and hence increasing $k_\parallel \rho_{e0}$. Secondly, the fraction of power at larger wavenumbers is significantly higher in 1D3V than in 2D3V. Indeed, in 2D3V the power in the higher wavenumbers actually decreases in time initially, while it increases in 1D3V, suggesting that energy transfer from smaller towards larger wavenumbers is easier in 1D3V. Conversely, the proportion of power in small- wavenumber modes is much larger in the 2D3V simulations. This suggests that in 2D, large-wavenumber modes are suppressed and more energy is retained in,  or transferred to, small-wavenumber modes. A natural interpretation is that this is due to the presence of oblique modes, which cannot form in 1D3V. The mechanism responsible for this redistribution is currently unclear and lies outside the scope of this paper. Finally, figure \ref{fig:timespectra} shows that the height of the spectral peak is similar for the same $\beta_{e0}$, regardless of whether the simulation was run in 1D3V or 2D3V. This suggests that the continued growth of whistler waves beyond $\delta B \sim B_0$ in 1D3V is due to these high wavenumber modes, and that the peak wavenumber saturates with a similar power regardless of the simulation dimensionality.

To further investigate the nature of the nonlinear wave-particle interactions and how they differ between 1D3V and 2D3V simulations, we characterise the magnetic moment of particles during reflection events. For this characterisation, the velocities parallel to the background magnetic field of individual particles present in 1D3V and 2D3V simulations with $\beta_{e0}=120$ were smoothed using a Savitsky-Golay filter \citep{savitzky_smoothing_1964} with window length corresponding to $10 \Omega_{e0}^{-1}$ to average out any Larmor motion. We then define a reflection event as the relative change in parallel velocity $\Delta v_\parallel / v_{\parallel 0} < -1$. The velocity of a particle between two reflection events $v_{\parallel 0}$ was calculated by finding the extremum of the smoothed velocity between the two changes in sign, and averaging $v_{\parallel}$ over $10 \Omega_{e0}^{-1}$ around that point. The change in velocity $\Delta v_{\parallel}$ was calculated by finding the difference between two such consecutive values. The magnetic moment between two reflection events was calculated by taking the average magnetic moment $\mu = m_e v_\perp^2 / 2B$ over the same time interval as the velocity average, and the difference $\Delta \mu$ was calculated in the same way.

We show the results of this analysis in figure \ref{fig:mag_moment}, where the relative changes in magnetic moment and parallel velocity during reflection events are plotted. The magnetic moment is approximately conserved in 1D3V simulations, while it is not in 2D3V simulations. There are also clearly two distinct populations in the 1D3V figure, due to the different Doppler shifts experienced by electrons travelling towards the hot and cold wall, respectively, interacting with whistler waves. These plots suggest that, in 1D3V, large-reflection events conserve $\mu$ well, consistent with magnetic-mirror-like reflection, whereas 2D3V simulations include other interactions, such as resonant scattering, which do not conserve $\mu$. As such, we conclude that the wave-particle interactions in 1D3V are distinct from those in 2D3V. This is, perhaps, unsurprising, as previous calculations have shown that resonant scattering of the heat-flux-carrying electrons cannot occur in 1D3V, whereas it can in 2D3V; however, this observation could explain the differences observed between the whistler dispersion relations for 1D3V and 2D3V simulations.

The fact that the advective heat flux theory describes both 2D3V and 1D3V simulations raises the possibility that 1D3V simulations could be used as a reduced model for investigating the WHFI in ultra-high-$\beta$ plasmas. This would substantially reduce the computational cost and make it possible to explore plasmas at higher $\beta_{e0}$ compared to what can plausibly be investigated in 2D3V simulations. However, the quantitative behaviour in 1D3V differs from that in 2D3V; $\delta B^2 / B_0^2$ continues to increase beyond $\delta B \sim B_0$, and the heat flux is less strongly suppressed. Thus, 2D3V simulations remain necessary for quantitatively reliable predictions of the WHFI in physical plasmas. Nevertheless, ultra-high-$\beta$ 1D3V simulations may be useful for quickly identifying parameter regimes in which interesting changes in behaviour occur, before more physically complete 2D3V simulations are performed.

\subsection{Physical implications and applications}\label{sec:disc:physical_implications}

Perhaps the most important implication of the present work is that, despite the large electron mean free path, heat transport in the ultra-high-$\beta$ regime appears to become effectively local once the WHFI saturates. Consequently, the advective heat-transport model supported by our simulations has a simple form that could be implemented in magneto-hydrodynamic (MHD) codes to better model collisionless, ultra-high-$\beta$ plasmas. Indeed, the scaling relation found for parallel heat transport $q_{e\parallel} \approx 4.7\beta_{e0}^{-1} q_{\mathrm{fs}}$ (where the prefactor $4.7$ is the mean prefactor from the 2D3V simulations with angled and parallel magnetic fields) is surprisingly close to that found in \citet{Komarov_Schekochihin_Churazov_Spitkovsky_2018}: $q_{e\parallel} \approx 1.5 \beta_{e0}^{-1} q_0 \approx 3 \beta_{e0}^{-1} n_e v_{\mathrm{th}e} T_e \approx 5.3\beta_{e0}^{-1} q_{\mathrm{fs}}$, despite the fact that the amplitude of the whistler waves becomes independent of $\beta_{e0}$ in the ultra-high-$\beta$ regime. Therefore, the same heat transport scaling with $\beta_{e0}$ seems to be observed in all WHFI-unstable collisionless high-$\beta$ regimes studied to date, regardless of the amplitude of the waves. This is a non-trivial result, as the very different mechanisms at play do not imply that this should be the case. For a typical front-side blow-off laser plasma with $T_e \sim 1 \mathrm{keV}$ and $\beta_{e0}=100$, the electron Larmor period is of the order $\Omega_e^{-1} \sim 1\times 10^{-13} \mathrm{s}$. This paper has shown that the saturation time for the WHFI is $\sim 10^3 \Omega_e^{-1}$, which corresponds to $\sim$0.1ns in the laser plasma; however, the heat flux itself approaches a constant value more quickly, after just several hundred electron Larmor periods. This is short compared to the time-scale for macroscopic evolution of the plasma, which is on the order of $\sim$1ns. It should therefore be possible to implement a heat flux model into a MHD code that follows $q_{e\parallel} \approx 4.7 \beta_{e0}^{-1} q_\mathrm{fs}$ for regions of the plasma where the WHFI operates (i.e. $\beta_e \lambda_e / L_\mathrm{T} \gg 1$). The locality of the heat flux in this model makes it much more computationally manageable compared to current non-local methods (e.g. \citet{luciani_nonlocal_1983, schurtz_nonlocal_2000}).

ICF plasmas are known to self-generate magnetic fields of order $10$-$100$ MG, enough to render the plasma weakly magnetised with a Hall parameter greater than unity \citep{walsh_self-generated_2017, walsh_biermann_2021, frank_self-generated_2024,frank_modeling_2025}. Prior research has shown that even weakly collisional plasmas are kinetically unstable for high enough $\beta_{e0}$ \citep{bott_kinetic_2024,lopez_collisional_2025}; hence the WHFI is likely to be dynamically important in ICF plasmas. Understanding how this instability impacts the transport properties of ICF plasmas is therefore vital for accurate modelling and optimising experimental designs. For example, \citet{walsh_magnetized_2025} showed that, for pre-magnetised ICF implosions simulated with the classical Braginskii heat conduction model, the best-performing capsule design was distinct from the unmagnetised case, and ignition could be achieved at reduced drive energies. How WHFI-regulated heat transport affects the performance of different capsule designs remains to be seen. If the present scaling continues to hold for ICF-hotspot relevant parameters ($\beta_e \sim 10^{3}$-$10^4$), we would predict suppression of heat fluxes by up to an order of magnitude, suggesting that exploiting this instability could, in principle, improve hot-spot energy confinement and thereby increase neutron yield.  

Outside of ICF research, these results may also be applicable to studies of the early Universe in the collisionless, ultra-high-$\beta$ plasma of the post re-ionisation inter-galactic medium. Much research has been carried out in primordial plasmas to investigate different sources that may have contributed to the growth of magnetic fields observed in present-day galaxies \citep{turner_inflation-produced_1988, vachaspati_magnetic_1991, ratra_cosmological_1992, widrow_origin_2002, de_souza_origin_2008, durrer_cosmological_2013, subramanian_origin_2016, vachaspati_progress_2021, di_valentino_primordial_2024}. These plasmas are thought to be ultra-high-$\beta$ and weakly collisional, with values of $\beta_e$ as high as $\beta_e \sim 10^{22}$ being possible~\citep{barkana_beginning_2001, bott_kinetic_2024}. As such, kinetic instabilities are likely to have been present, with potentially significant impacts on both transport properties and growth of the magnetic energy. In particular, this paper has shown that kinetic instabilities like the WHFI can significantly amplify magnetic energy. The 2D3V simulations showed significant magnetic field energy at scales significantly larger than the local electron Larmor radius, raising the possibility that the WHFI may act as an intermediary for transferring magnetic energy from electron to ion scales, before ion kinetic instabilities amplify the magnetic fields further.

Finally, the results of this paper have implications for laboratory-astrophysics experiments investigating anomalous heat conduction in weakly collisional plasmas~\citep[e.g.,][]{vincent_design_2026}. In order for the results of such experiments to be interpreted correctly, it is important to understand whether the scaling relationships in the laboratory plasma and its astrophysical counterpart are the same. As this paper has shown, the transport mechanisms differs between plasmas with $\beta_{e0} \rho_{e0} / L_{\mathrm{T}0} \gtrsim 1$ and those with $\beta_{e0} \rho_{e0} / L_{\mathrm{T}0} \ll 1$; hence any laboratory experiments investigating heat transport in astrophysical plasmas must ensure this parameter is in the appropriate regime for the phenomena they are studying.

\subsection{Limitations and future work} \label{sec:disc:lims}

While the present work establishes a theoretical framework for the WHFI and its regulation of heat transport in ultra-high-$\beta$ plasmas, and demonstrates good agreement with kinetic simulations, several important questions remain unanswered. First, the simulations presented here are collisionless, which limits their direct applicability of the results to weakly collisional plasmas. Extending the present work to weakly collisional plasmas is therefore a natural next step. Previous research has shown that collisions damp, but do not necessarily prevent the formation of, kinetic instabilities \citep[see][]{lopez_collisional_2025}. It is therefore plausible that, in a weakly collisional plasma, the whistler waves would not attain the same amplitudes that they do in the collisionless case, and that the diffusive heat flux may become more significant due to the presence of collisions, although this remains to be investigated.

A second limitation is that the simulations presented here are 1D3V and 2D3V. Although the 2D3V simulations allow both parallel and oblique whistler modes, and can investigate cross-field transport in the simulation plane, they do not capture fully three-dimensional (3D) mode coupling or magnetic-field-line wandering. The latter effect might be expected to be important in ultra-high-$\beta$ plasmas because the magnetic-field perturbations associated with the whistler fluctuations are comparable in size to the background magnetic field. Extending the present work to fully 3D simulations is therefore important, though we caution that such simulations using current kinetic codes would have a high computational cost.  

Another limitation is that we do not propose a complete theory that explains all of the quantitative properties of the WHFI in ultra-high-$\beta$ plasmas that we observe in our simulations. For example, the exact process by which the whistler waves acquire a linear dispersion relation in the ultra-high-$\beta$ regime is not studied here. Likewise, the physical origin of the difference between the dispersion relation in 1D3V simulations compared with 2D3V simulations is also not fully understood. This presents an interesting avenue for future research into nonlinear wave-wave and wave-particle interactions in high-$\beta$ plasmas.  

Another key question that remains unanswered is the upper limit on $\beta_e$ at which the WHFI no longer regulates thermal transport. As unmagnetised plasmas are not susceptible to the WHFI, it should be possible to derive a bound for $\beta_e$, above which the heat flux is no longer suppressed as $\beta_e$ increases further. In a collisionless plasma, it seems reasonable to assume that the WHFI can no longer operate when the Larmor radius becomes a finite fraction of the temperature length scale due to the insufficient scale separation between the characteristic whistler wavelength $2\pi / k_{\rm ww} \approx 2\pi / \rho_e$ and $L_\mathrm{T}$. This implies that an upper bound on $\beta_e$ to destabilise the WHFI would be $\beta_{e} \sim L_\mathrm{T}^2/d_e^2$, where $d_e = \beta_e^{-1/2} \rho_e$ is the electron skin depth. Investigating the exact nature of this upper bound for $\beta_e$ is left to future work.

The WHFI, and kinetic instabilities more generally, remain comparatively unexplored in ultra-high-$\beta$ plasmas, despite its relevance to HED plasmas, laboratory astrophysics, and the early Universe. In this paper, we have demonstrated that large-amplitude whistler fluctuations fundamentally alter the mechanisms supporting electron heat transport, replacing the quasilinear pitch of resonant pitch-angle scattering with one in which thermal energy is regulated by the propagation of large-amplitude magnetic fluctuations that act as effective transport barriers. Understanding the nonlinear plasma physics responsible for this transition, and understanding how robust it remains in weakly collisional plasmas, is essential for understanding and developing predictive heat-flux models in high-$\beta$ plasmas countered in both nature and the laboratory. 

\section{Acknowledgements}
RD was supported by the EPSRC and First Light Fusion under the AMPLIFI Prosperity Partnership - EP/X025373/1; PPC and AFAB were supported by UKRI (grant number MR/W006723/1). Computing resources provided by STFC Scientific Computing Department’s SCARF cluster. The authors would like to thank Adam Fraser and Alex Schekochihin for helpful and engaging discussions on this work.

\bibliographystyle{jpp}

\bibliography{bibliography}

\end{document}